\PassOptionsToPackage{dvipsnames}{xcolor}
\PassOptionsToPackage{normalem}{ulem}
\documentclass{bytedance_seed}

\usepackage[utf8]{inputenc}
\usepackage{amsfonts}
\usepackage{nicefrac}
\usepackage{amsmath}
\usepackage{amssymb}
\usepackage{mathtools}
\usepackage{amsthm}
\usepackage{algorithm}
\usepackage{algorithmic}
\usepackage{wrapfig}
\usepackage{colortbl}
\usepackage{float}
\usepackage{listingsutf8}
\usepackage{pifont}
\microtypesetup{expansion=false}

\theoremstyle{plain}

\theoremstyle{definition}

\theoremstyle{remark}

\title{\textsc{SWE-MeM}: Learning Adaptive Memory Management for \\ Long-Horizon Coding Agents}

\newtcblisting{PromptBox}{
  breakable,
  enhanced,
  listing only,
  width=\linewidth,
  listing engine=listings,
  colback=gray!4,
  colframe=gray!60,
  boxrule=0.5pt,
  arc=1.2mm,
  left=0.6mm,
  right=0.6mm,
  top=0.8mm,
  bottom=0.8mm,
  listing options={
    basicstyle=\tiny\ttfamily,
    breaklines=true,
    breakatwhitespace=false,
    breakautoindent=false,
    breakindent=0pt,
    columns=fullflexible,
    keepspaces=false,
    showstringspaces=false,
    upquote=true,
    tabsize=2,
    basewidth=0.46em,
    inputencoding=utf8,
    extendedchars=true,
    literate={→}{{$\rightarrow$}}1 {—}{{--}}1
  }
}

\author[1]{Shuzheng Gao}
\author[2]{Wenhao Zeng}
\author[3]{Zhaojian Yu}
\author[4]{Jianqiao Wangni}
\author[1]{Chaozheng Wang}
\author[4]{Kai Cai}
\author[4,\dagger]{Shilin He}
\author[1]{Michael R. Lyu}

\affiliation[1]{The Chinese University of Hong Kong, China}
\affiliation[2]{Shanghai Jiao Tong University, China}
\affiliation[3]{Tsinghua University, China}
\affiliation[4]{ByteDance, China}

\contribution[\dagger]{Corresponding author}
\correspondence{Shilin He, \email{heshilin@bytedance.com}}

\abstract{
Long-horizon software engineering agents often need to manage lengthy and noisy interaction histories under limited context budgets. Existing memory management methods typically rely on static compression workflows or impose rigid constraints on compression timing and granularity.
Moreover, these approaches fail to jointly optimize memory management and issue resolution capabilities to improve performance while reducing token usage. We present \textsc{SWE-MeM}, a training framework for proactive and on-demand memory management in software engineering agents. \textsc{SWE-MeM} provides a flexible memory tool that lets agents decide when, what, and how to compress based on trajectory state, task progress, and remaining context budget. We train agents with synthesized proactive memory-management trajectories and Memory-aware GRPO, which jointly optimizes memory management and issue resolution through memory-aware trajectory splitting and step-level credit assignment. On SWE-Bench Verified, \textsc{SWE-MeM} achieves 43.4\% and 60.2\% resolve rate with 4B and 30B models, respectively, outperforming existing memory management baselines in both performance and efficiency.
}

\date{\today}

\fancypagestyle{firststyle}{%
  \fancyhf{}%
  \fancyhead[L]{\includegraphics[height=10mm]{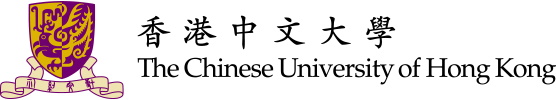}}%
  \fancyhead[R]{\includegraphics[height=10mm]{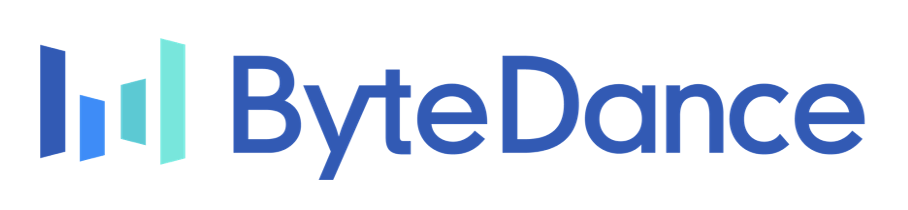}}%
}
\makeatletter
\def\fps@figure{htbp}
\def\fps@table{htbp}
\makeatother

\begin{document}
\maketitle

\section{Introduction}

Solving long-horizon software engineering tasks remains a key challenge for deploying agents in real-world development workflows. Most existing agents follow a ReAct-style append-only paradigm~\cite{ReAct2023,SWEAgent2024,OpenHands2025}, where interaction history is continuously accumulated to model context over time. In software engineering tasks, numerous intermediate actions and lengthy environment feedback can quickly accumulate into long and noisy trajectories, often exceeding the context window of LLMs~\cite{DBLP:journals/corr/abs-2511-13998,DBLP:journals/corr/abs-2604-22750}. Even when the history fits within the window, agents still face the challenge of effectively utilizing task-relevant context information from long and noisy trajectories~\cite{DBLP:journals/corr/abs-2602-03587,DBLP:conf/emnlp/DuTRRBGWSHP25}. These challenges make effective memory management essential for long-horizon coding agents.

To address these challenges, recent work has explored memory management through threshold-based summarization, subtask-level folding, and learned compression tools~\cite{DBLP:journals/corr/abs-2509-13313,DBLP:journals/corr/abs-2510-11967,DBLP:journals/corr/abs-2512-22087}. These methods reduce context length by summarizing the full history once a threshold is reached~\cite{DBLP:journals/corr/abs-2509-13313}, folding completed subtasks~\cite{DBLP:journals/corr/abs-2510-11967}, or compressing earlier conversation turns while preserving recent interactions~\cite{DBLP:journals/corr/abs-2512-22087}.

\begin{figure}[t]
    \centering
    \begin{minipage}[t]{0.49\linewidth}
        \centering
        \includegraphics[width=\linewidth,trim={0 500 0 0},clip]{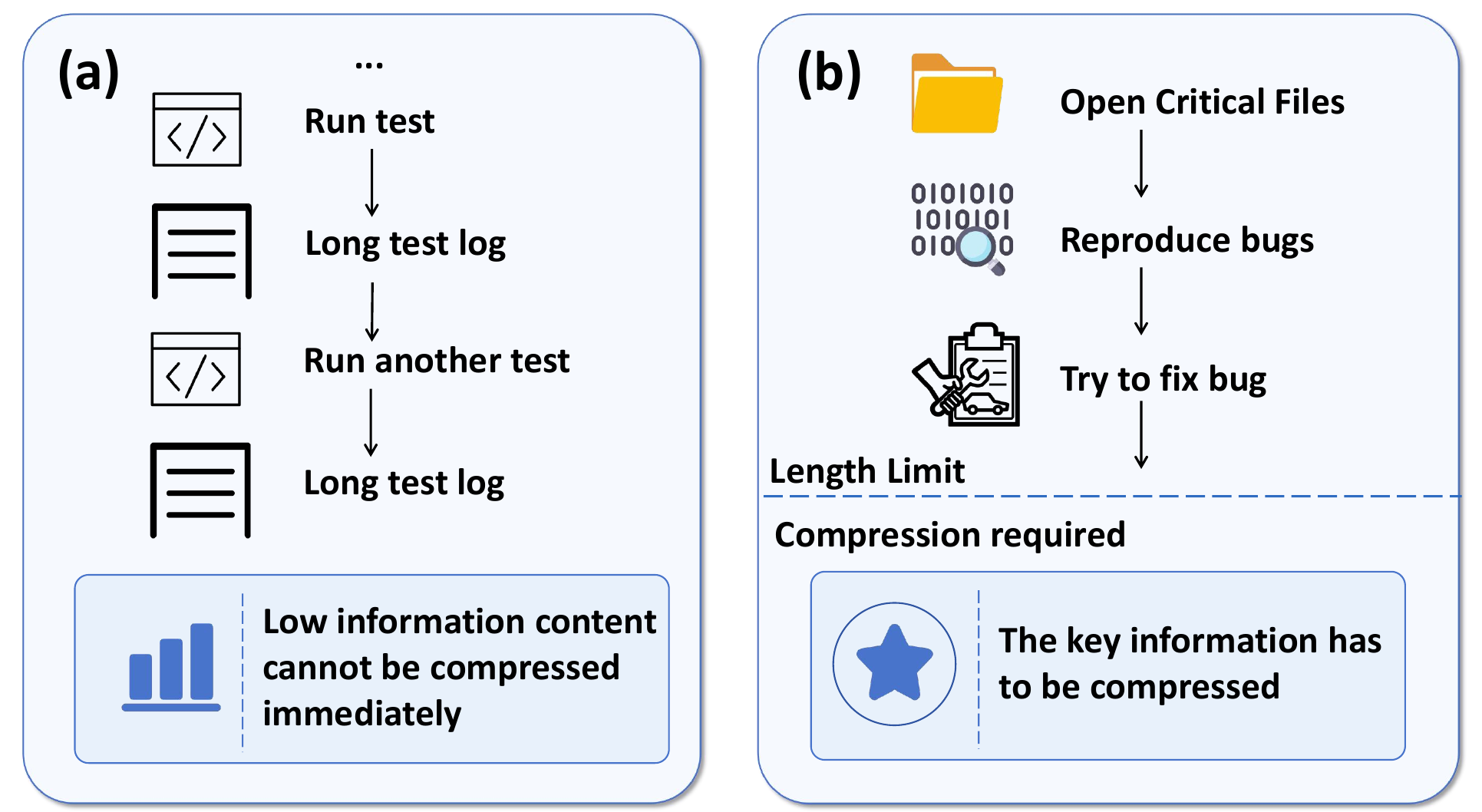}
    \end{minipage}\hfill
    \begin{minipage}[t]{0.51\linewidth}
        \centering
        \includegraphics[width=\linewidth,trim={0 0 0 508},clip]{figures/limitation_revised.pdf}
    \end{minipage}
    \caption{Illustration of the tool flexibility issue in (a) and (b), and the performance--cost comparison in (c), where lower-right is better. \textsc{SWE-MeM} achieves the highest resolve rate while maintaining relatively low token and step usage compared with other methods.}
    \label{fig:intro_overview}
\end{figure}

However, they remain limited in both flexibility and optimization objective. First, fixed compression rules cannot adapt to the heterogeneous value of information in software engineering trajectories: lengthy execution logs or irrelevant files can typically be safely compressed at any point, whereas code inspected early in the trajectory may remain essential for later reasoning and therefore may not be suitable for compression, as illustrated in Figure~\ref{fig:intro_overview}(a) and (b). Second, existing methods typically optimize context reduction rather than task-solving efficiency. As shown in Figure~\ref{fig:intro_overview}(c), although existing methods can substantially reduce the agent’s peak context length, they still increase total token usage or interaction steps dramatically, leading to worse overall efficiency.

To address these issues, we propose \textsc{SWE-MeM}, a training framework that equips long-horizon software engineering agents with proactive and on-demand memory management. \textsc{SWE-MeM} provides a flexible memory tool that allows the model to decide when to compress, what to compress, and how to summarize the selected content based on the current trajectory state. We train this capability by synthesizing task-solving trajectories with proactive memory-tool invocations, where the compression timing and targets are chosen according to the current trajectory state and remaining subtasks. We further apply curriculum learning to teach effective tool use, especially in proactive compression scenarios. Finally, we introduce Memory-aware GRPO, which jointly improves issue resolution and memory management through memory-aware trajectory splitting and step-level credit assignment. On SWE-Bench Verified, \textsc{SWE-MeM} achieves resolve rates of 43.4\% and 60.2\% with Qwen3-4B-Instruct and Qwen3-Coder-30B-A3B, respectively. Under a 32K context budget, it outperforms existing memory management methods with higher resolve rates and fewer interaction rounds. In summary, we make the following contributions:

    
    

\begin{itemize}
    \item We introduce an adaptive memory tool that enables agents to decide when to compress, what to compress, and how to summarize the selected context based on the current trajectory state.

    \item We propose a training framework that combines memory-management trajectory synthesis, curriculum fine-tuning, and Memory-aware GRPO to jointly optimize memory decisions and task-solving performance.

    \item Experiments on SWE-Bench Verified show that \textsc{SWE-MeM} achieves resolve rates of 43.4\% with a 4B model and 60.2\% with a 30B model under a 32K context budget, outperforming existing memory management methods in both performance and efficiency.
\end{itemize}




\FloatBarrier

\section{Proposed Approach}\label{sec:method}

\begin{figure*}[t]
    \centering
    \includegraphics[width=1\linewidth]{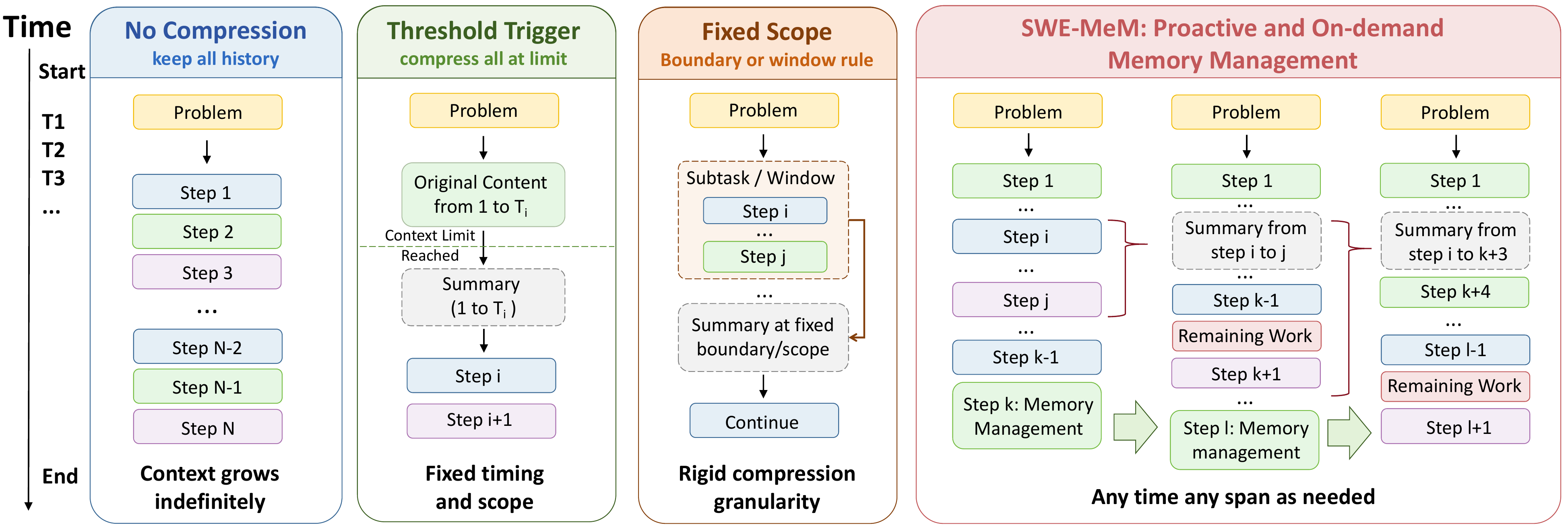}
    \caption{Comparison of different memory management methods. No compression mode causes the context to grow indefinitely; threshold-trigger methods compress only after the budget is reached; fixed-scope methods rely on rigid boundaries or windows; in contrast, \textsc{SWE-MeM} performs proactive and on-demand memory management over flexible context spans. 
    }
    \label{fig:overview}
\end{figure*}

\subsection{Memory Management Tool Design}

We introduce a flexible memory management tool that enables software engineering agents to decide when to invoke compression, which span to compress, and how to summarize the selected span, rather than following fixed compression workflows. To formalize the process, we represent the interaction history at time step $t$ as:
\begin{equation}
\begin{aligned}
C_t = \big\langle
&p, \{a_1, o_1\}, \{a_2, o_2\}, \ldots, \\
&\{a_{t-1}, o_{t-1}\}
\big\rangle,
\end{aligned}
\end{equation}
where $p$ is the initial problem statement, $a_i$ denotes the agent action at step $i$, including textual responses, reasoning traces, and tool calls, and $o_i$ denotes the corresponding environment observation, augmented with metadata such as the current step number and remaining context length to inform the agent of its progress and context budget.

When the agent determines that compression is required, it can invoke this tool
\texttt{compress(analysis, start\_\allowbreak{}step\_\allowbreak{}number,
end\_\allowbreak{}step\_\allowbreak{}number, content , remaining\_\allowbreak{}work)} to proactively manage its context.
Here, \texttt{analysis} records the agent's assessment of the current progress and unresolved subtasks. The arguments \texttt{start\_step\_number} $s$ and \texttt{end\_step\_number} $e$ specify the span to compress. The selected span usually contains raw details that are less necessary for subsequent reasoning such as lengthy logs, failed exploratory attempts, or obsolete intermediate observations. The \texttt{content} argument $c$ is the summary that replaces the selected span, which distills the original information into a higher-level abstraction by removing unnecessary details. Finally, \texttt{remaining\_work} $f$ records the remaining subtasks and is appended to the end of the trajectory to maintain task continuity. The agent generates all arguments in this order within a single LLM call. As shown in Figure~\ref{fig:overview}, the resulting compression operation is:
\begin{align}\label{equ:compression}
C_t \xrightarrow{\mathcal{M}} C'_t
&= \big\langle p, \{a_1, o_1\}, \ldots, \{a_{s-1}, o_{s-1}\}, \nonumber \\
&\quad \{c\}, \{a_{e+1}, o_{e+1}\}, \ldots, \nonumber \\
&\quad \{a_{t-1}, o_{t-1}\}, \{f\} \big\rangle .
\end{align}
Here, $\{c\}$ and $\{f\}$ are memory entries inserted into the trajectory. 
This operation preserves temporal ordering while reducing context length. Since a compressed summary can itself be selected in later tool calls, the agent can progressively compress the content as the trajectory grows.

\begin{figure*}[t]
    \centering
    \includegraphics[width=1\linewidth]{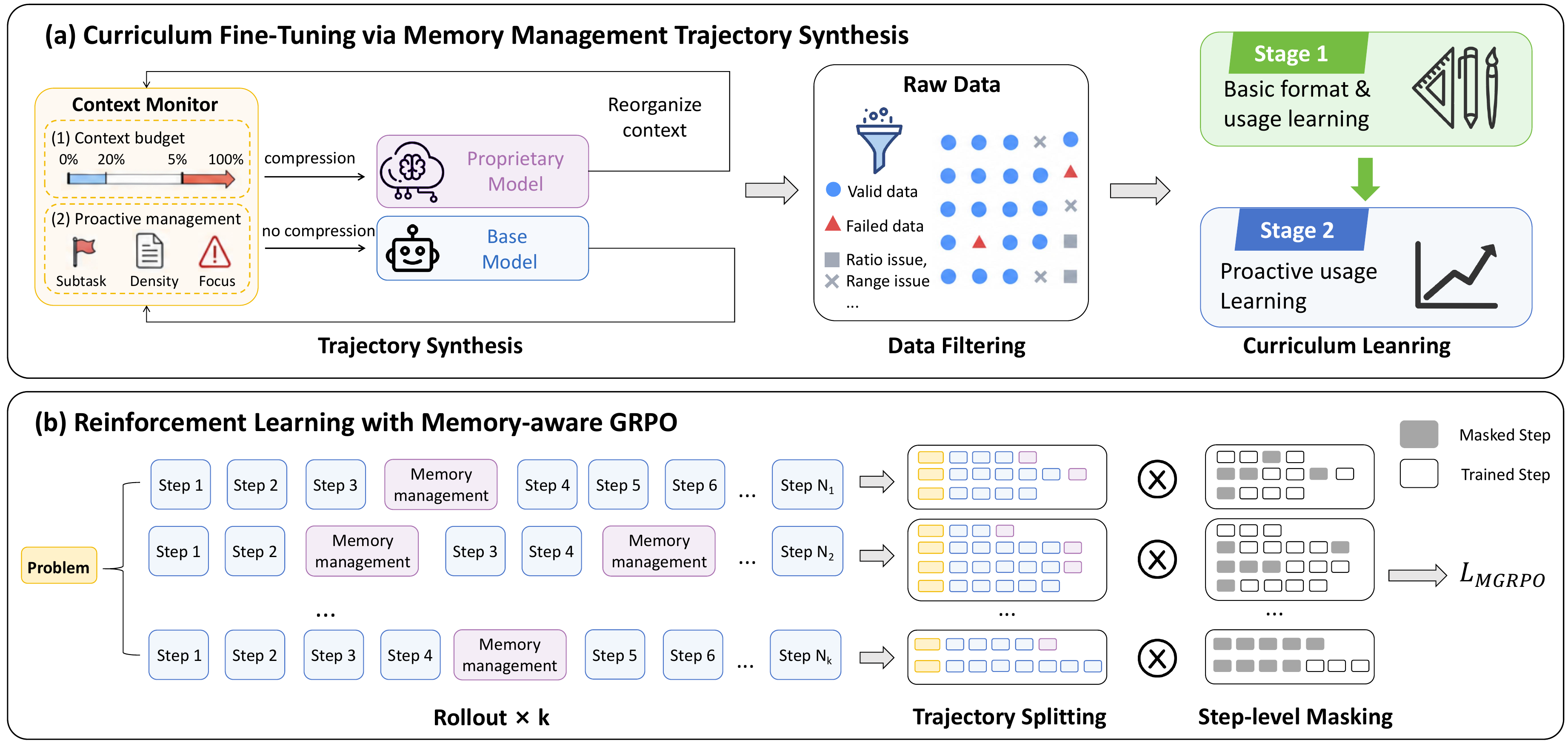}
    \caption{Overview of the \textsc{SWE-MeM} training pipeline.}
    \label{fig:method}
\end{figure*}

\subsection{Curriculum Fine-Tuning with Synthesized Memory Trajectories}

To bootstrap the model's ability to use the memory tool, we develop a synthesis workflow that augments task-solving trajectories with memory-management actions, together with a curriculum training method for learning proactive memory behavior, as shown in Figure~\ref{fig:method} (a).

\subsubsection{Trajectory Synthesis Workflow}

The synthesis workflow integrates task solving and memory management into a single trajectory-generation process, as detailed in Algorithm~\ref{alg:trajectory-synthesis} in Appendix~\ref{sec:method_detail}. For each task, the rollout is generated by the base model that is later fine-tuned, rather than by a stronger teacher model. The proprietary model is used only for proactive-trigger judgment and memory-tool argument synthesis; it does not generate task-solving actions.

As shown in Figure~\ref{fig:method}(a), we introduce a context monitor that checks whether a memory-management action should be triggered at the beginning of each step. Inspired by existing agent context-compression designs~\cite{DBLP:journals/corr/abs-2510-11967,DBLP:journals/corr/abs-2509-13313,DBLP:journals/corr/abs-2509-23586}, the monitor considers two types of triggers that reflect different trajectory states: context budget pressure and proactive memory management.

\textbf{Context budget pressure} addresses imminent context exhaustion with a probabilistic scheduler. The trigger probability becomes non-zero when the remaining context budget drops below 20\% and increases monotonically as the budget shrinks, reaching 1 when only 5\% of the budget remains. This soft schedule avoids relying on a single hard threshold and improves the diversity of synthesized trajectories.

\textbf{Proactive memory management} captures situations where compression is useful before the context budget becomes critical. We consider three common cases: \textit{subtask completion}, where a subtask has been completed and only its conclusion needs to be retained; \textit{low-information density}, where recent rounds are dominated by verbose logs, irrelevant file exploration, or other low-value content; and \textit{focus degradation}, where long and noisy context appears to distract the agent from the core issue, and compression can restore a cleaner working context. These cases are detected with an LLM-as-a-judge mechanism, with detailed prompts provided in Appendix~\ref{sec:prompt}.

Once a trigger fires, we invoke the proprietary model to synthesize the memory-tool arguments, and the trajectory is reorganized according to Eq.~\ref{equ:compression}. The synthesis loop continues until the agent resolves the issue or reaches the maximum number of steps.



\subsubsection{Trajectory Quality Filtering}\label{subsec:fitering}

We first perform rejection sampling to retain only trajectories that pass the test cases. We then apply a rule-based data-cleaning pipeline to reduce noise in the retained trajectories before training. Rather than discarding an entire trajectory when a local defect is detected, we mask the corresponding assistant message during training, allowing the remaining valid steps to contribute to learning.

Our filtering criteria cover both compression and non-compression actions. For compression actions, we filter three types of defects. \textbf{(1) Inappropriate reduction ratio} refers to summaries that are either too aggressive or too conservative. We define the reduction ratio as the fraction of tokens removed from the selected span: ratios above 80\% may discard critical information, whereas ratios below 20\% provide little practical context saving. \textbf{(2) Overly short compression ranges} occur when the selected span is too short to yield meaningful context reduction, making the compression overhead unnecessary. \textbf{(3) Range misalignment} occurs when the summary does not faithfully correspond to the declared step range, either omitting information from steps inside the range or incorporating content from outside the range. For non-compression actions, \textbf{(4) low-quality responses} are masked if they contain empty outputs, repetitions, truncations, or invalid tool calls.

\subsubsection{Curriculum Learning for Proactive Memory Management}

Proactive memory management is harder to learn than budget-based compression because it depends on semantic judgments about trajectory state. 
In our synthesized trajectories, budget-based compression also appears more frequently and follows clearer context-budget patterns. We therefore adopt a two-stage curriculum learning strategy to progressively train the model to use the memory tool, with a particular focus on proactive memory behavior.

In the first stage, we train the model on the full set of filtered synthetic trajectories, allowing it to learn the basic format and usage patterns of memory-management tools. In the second stage, we construct a more targeted training set. We first extract trajectories that contain proactive memory-management actions and apply rubric-based filtering to retain high-quality instances. The rubrics check whether the \texttt{analysis} argument provides a reasonable assessment of the current progress and remaining subtasks, and whether the compressed content is consistent with this analysis. The detailed rubric prompts are provided in Appendix~\ref{prompt:stage2-compression-quality}.

To avoid catastrophic forgetting, we also sample an equal number of problems and collect trajectories that contain budget-based compression but no proactive memory-management actions. For this subset, we apply another set of LLM-as-a-judge rubrics to retain trajectories in which the agent effectively uses the compressed information in subsequent steps, avoids redundantly re-acquiring information already preserved in memory, and maintains coherent progress along the established solution path. Finally, we merge the two retained subsets for the second-stage fine-tuning, strengthening proactive memory behavior while preserving budget-based compression and general task-solving capabilities.

\subsection{Reinforcement Learning with Memory-aware GRPO}

To further improve memory-tool use and task-solving performance, we propose Memory-aware GRPO, which trains the model with reinforcement learning (RL) while accounting for the state changes introduced by compression.

\subsubsection{Memory-aware Trajectory Splitting and Loss Aggregation}

Unlike standard agent training, memory management changes the context state after each compression and therefore affects all subsequent actions. To maintain consistency with inference-time states, as shown in Figure~\ref{fig:method}(b), we checkpoint the trajectory at compression steps and split each rollout into sub-trajectories. Each sub-trajectory starts from the exact compressed prefix that the agent would observe online, rather than from the original uncompressed history. We apply the same splitting scheme during supervised fine-tuning (SFT) to maintain training-inference consistency.

Memory management can introduce large variance in trajectory length. Longer rollouts often contain inefficient memory operations or fail to reuse conclusions already preserved in compressed memory, leading to redundant actions. To reduce optimization bias from length variability, we aggregate the loss in two stages: we first average all valid token-level objectives within each rollout after splitting, and then average the resulting rollout-level losses across the batch:
\begin{equation}
\mathcal{L}
=
\frac{1}{S}
\sum_{i=1}^{S}
\bar{\ell}_i,
\quad
\bar{\ell}_i =
\frac{1}{N_i}
\sum_{k=1}^{K_i}\sum_{t=1}^{T_{i,k}}
\ell_{i,k,t}.
\end{equation}
Here, $\ell_{i,k,t}$ denotes the loss of token $t$ in sub-trajectory $k$ of rollout $i$, $N_i=\sum_{k=1}^{K_i}T_{i,k}$ is the number of valid tokens in rollout $i$, and $S$ is the number of rollout samples in the batch. This aggregation makes each rollout contribute equally to the final loss, preventing extremely long trajectories or sub-trajectories from being overweighted.

\subsubsection{Step-level Mask for Credit Assignment}

Standard GRPO assigns credit using only the final issue-resolution outcome, which may incorrectly penalize appropriate intermediate actions in failed trajectories or reward problematic actions in successful trajectories. We therefore introduce step-level credit masks on top of multi-trajectory GRPO. The step-level mask is broadcast to all tokens belonging to the corresponding assistant step:
\begin{equation}
\begin{aligned}
\mathcal{J}_{\mathrm{MGRPO}}
&= \mathbb{E}_{i}\!\Bigg[
\frac{1}{M_i}\sum_{k=1}^{K_i}\sum_{t=1}^{T_{i,k}} m_{i,k,t}
\times \min\!\Big(
r_{i,k,t}\hat{A}_i, \\
&\qquad
\mathrm{clip}(r_{i,k,t},1-\epsilon,1+\epsilon)\hat{A}_i
\Big)
\Bigg],
\end{aligned}
\end{equation}
where $m_{i,k,t}\in\{0,1\}$ controls whether token $t$ contributes gradient, and $M_i=\sum_{k=1}^{K_i}\sum_{t=1}^{T_{i,k}}m_{i,k,t}$ is the number of unmasked valid tokens in rollout $i$. We design the following masking strategies to improve credit assignment:

\textbf{Memory Management Action Quality.}
For memory-management actions in successful trajectories, we apply rule-based checks to mask steps with inappropriate reduction ratios, overly short compression ranges, or range misalignment.

\textbf{Late Memory Management.}
In successful trajectories, we also mask memory-management actions executed only after the remaining context budget has fallen below threshold $\tau$. Although these actions may not cause failure in the current trajectory, they delay memory management and may increase overflow risk in similar future states.

\textbf{Overflow Failure.}
If a trajectory exceeds the length limit without invoking memory management tools, we keep only the last overflow sub-trajectory for penalty assignment. Within this sub-trajectory, penalties are assigned only to steps where the remaining budget is below threshold $\tau$ but the agent still fails to invoke memory management.

\textbf{Invalid Tool Usage.}
For successful trajectories, we mask steps that contain invalid tool calls, such as calls with incorrect formats or arguments, or repetitive content, preventing the model from learning undesirable action patterns.

\FloatBarrier

\section{Experiment}\label{sec:setup}

\begin{table*}[t]
  \centering
  \resizebox{1\linewidth}{!}{%
    \begin{tabular}{lcccc}
      \toprule
      \textbf{Method/Model} & \textbf{Base Model} & \textbf{Base Scaffold} & \textbf{Length Limitation} & \begin{tabular}[c]{@{}c@{}}\textbf{Resolve}\\\textbf{Rate}\end{tabular} \\
      \midrule
      \rowcolor{gray!35}
      \multicolumn{5}{l}{\textbf{\quad <10B Models}} \\
      \rowcolor{gray!15}
      \multicolumn{5}{c}{\textbf{ReAct Agent}} \\
      Qwen3-4B-Instruct~\cite{DBLP:journals/corr/abs-2505-09388} & -- & OpenHands & 128k & 7.0 \\
      Qwen3-4B-Instruct~\cite{DBLP:journals/corr/abs-2505-09388} & -- & OpenHands & 32k & 5.2 \\
      SWE-Gym-7B~\cite{SWEGym2025} & Qwen2.5-Coder-7B & OpenHands & 32k & 10.6 \\
      SWE-Mirror-7B~\cite{SWEMirror2025} & Qwen2.5-Coder-7B & MOpenHands & 128k & 22.8 \\
      SWE-Dev-7B~\cite{WangHWTD25} & Qwen2.5-Coder-7B & OpenHands & 160k & 23.4 \\
      SWE-Master-4B~\cite{DBLP:journals/corr/abs-2602-03411} & Qwen3-4B-Instruct & OpenHands & 128k & 33.4 \\
      Klear-Agent-8B~\cite{DBLP:journals/corr/abs-2511-05951} & Qwen3-8B & Mini-SWE-agent-plus & 64k & 39.4 \\
      SWE-Lego-8B~\cite{DBLP:journals/corr/abs-2601-01426} & Qwen3-8B & OpenHands & 128k & 42.2 \\
      \midrule
      \rowcolor{gray!15}
      \multicolumn{5}{c}{\textbf{\textsc{SWE-MeM}}} \\
      \textsc{SWE-MeM} SFT & Qwen3-4B-Instruct & OpenHands & 32k & 41.6 \\
      \rowcolor{green!6} \textsc{SWE-MeM} SFT+RL & Qwen3-4B-Instruct & OpenHands & 32k & 43.4 \\
      \midrule
      \rowcolor{gray!35}
      \multicolumn{5}{l}{\textbf{\quad \(\sim\)30B Models}} \\
      \rowcolor{gray!15}
      \multicolumn{5}{c}{\textbf{ReAct Agent}} \\
      Qwen3-Coder-30B-A3B~\cite{DBLP:journals/corr/abs-2505-09388} & -- & OpenHands & 256k & 51.6 \\
      Qwen3-Coder-30B-A3B~\cite{DBLP:journals/corr/abs-2505-09388} & -- & OpenHands & 32k & 38.6 \\
      Seed-OSS-36B~\cite{SeedOSS36BHF} & -- & OpenHands & 327K & 55.2 \\
      SWE-Gym-32B~\cite{SWEGym2025} & Qwen2.5-Coder-32B & OpenHands & 32k & 20.6 \\
      SWE-Dev-32B~\cite{WangHWTD25} & Qwen2.5-Coder-32B & OpenHands & 160k & 36.6 \\
      DeepSWE-32B~\cite{DeepSWE2025} & Qwen3-32B & R2E-Gym & 128k & 43.2 \\
      SWE-Mirror-32B~\cite{SWEMirror2025} & Qwen2.5-Coder-32B & MOpenHands & 128k & 52.2 \\
      SWE-Lego-32B~\cite{DBLP:journals/corr/abs-2601-01426} & Qwen3-32B & OpenHands & 128k & 52.6 \\
      \midrule
      \rowcolor{gray!15}
      \multicolumn{5}{c}{\textbf{Summary Agent}} \\
      Threshold-Compression Agent \cite{DBLP:journals/corr/abs-2512-22087} & Qwen2.5-Coder-32B & OpenHands & 64k & 53.8 \\
      Threshold-Compression Agent \cite{DBLP:journals/corr/abs-2510-11967} & Seed-OSS-36B & OpenHands & 32k & 55.0 \\
      \midrule
      \rowcolor{gray!15}
      \multicolumn{5}{c}{\textbf{Folding Agent}} \\
      SWE-Compressor (150 step)~\cite{DBLP:journals/corr/abs-2512-22087} & Qwen2.5-Coder-32B & OpenHands & 64k & 54.8 \\
      SWE-Compressor~\cite{DBLP:journals/corr/abs-2512-22087} & Qwen2.5-Coder-32B & OpenHands & 64k & 57.8 \\
      Context Folding~\cite{DBLP:journals/corr/abs-2510-11967} & Seed-OSS-36B & OpenHands & 32k & 58.0 \\
      \midrule
      \rowcolor{gray!15}
      \multicolumn{5}{c}{\textbf{\textsc{SWE-MeM}}} \\
      \textsc{SWE-MeM} Workflow-only & Qwen3-Coder-30B-A3B & OpenHands & 32k & 58.4 \\
      \textsc{SWE-MeM} SFT (150 step) & Qwen3-Coder-30B-A3B & OpenHands & 32k & 57.2 \\
      \textsc{SWE-MeM} SFT & Qwen3-Coder-30B-A3B & OpenHands & 32k & 58.8 \\
      \rowcolor{green!6} \textsc{SWE-MeM} SFT+RL & Qwen3-Coder-30B-A3B & OpenHands & 32k & 60.2 \\
      \bottomrule
    \end{tabular}%
  }
  \caption{Performance comparison on SWE-Bench Verified.
  }
  \label{tab:model_performance}
\end{table*}

\subsection{Datasets}
For training data, we construct trajectories from SWE-ReBench~\cite{SWERebench2025} and SWE-Gym~\cite{SWEGym2025}. For SFT, task-solving rollouts are generated by Qwen3-Coder-30B-A3B, while GPT-5.1 is used only for proactive-trigger judgment and memory-tool argument synthesis. For RL, we build filtered training groups from these two datasets by removing problems whose sampled rollouts are either all successful or all failed, retaining groups with non-trivial outcome variation. For evaluation, we mainly use SWE-Bench Verified~\cite{SweBench2024} and report resolve rate as the primary metric. Detailed dataset construction and evaluation settings are provided in Appendix~\ref{sec:appendix_dataset_details}.

\begin{table*}[tbp]
  \centering
  \resizebox{1\linewidth}{!}{%
    \begin{tabular}{lccccc}
      \toprule
      \textbf{Method} & \begin{tabular}[t]{@{}c@{}}\textbf{Resolve}\\\textbf{Rate}\end{tabular} & \textbf{Token Usage} & \begin{tabular}[t]{@{}c@{}}\textbf{Relative Token}\\\textbf{Usage}\end{tabular} & \textbf{Average Step} & \begin{tabular}[t]{@{}c@{}}\textbf{Relative Step}\\\textbf{Usage}\end{tabular} \\
      \midrule
      \rowcolor{gray!15}
      \multicolumn{6}{c}{\textbf{Summary Agent}} \\
      Threshold-Compression Agent \cite{DBLP:journals/corr/abs-2512-22087} & 53.8 & 5.18M & 203.9\% & -- & -- \\
      Threshold-Compression Agent \cite{DBLP:journals/corr/abs-2510-11967} & 55.0 & -- & -- & 74.9 & 151.3\% \\
      \midrule
      \rowcolor{gray!15}
      \multicolumn{6}{c}{\textbf{Folding Agent}} \\
      SWE-Compressor (150 step) & 54.8 & 1.89M & 96.4\% & -- & -- \\
      SWE-Compressor & 57.8 & 2.75M & 108.3\% & -- & -- \\
      Context Folding & 58.0 & -- & -- & 96.5 & 194.9\% \\
      \midrule
      \rowcolor{gray!15}
      \multicolumn{6}{c}{\textbf{\textsc{SWE-MeM}}} \\
      \textsc{SWE-MeM} SFT (150 step) & 57.2 & 0.86M  & 89.0\% & 73.8 & 119.6\% \\
      \textsc{SWE-MeM} SFT & 58.8 & 0.91M & 94.7\% & 77.7 & 125.9\% \\
      \rowcolor{green!6} \textsc{SWE-MeM} SFT+RL & 60.2 & 0.91M & 94.7\% & 77.0 & 123.5\% \\
      \bottomrule
    \end{tabular}%
  }
  \caption{Efficiency comparison with memory management baselines on SWE-Bench Verified. Relative usage compares token usage and the number of steps with those of the corresponding base models using the ReAct agent.}
  \label{tab:model_efficiency}
\end{table*}

\subsection{Baselines}

We compare our approach against three categories of baselines.

(1) \textbf{ReAct agents without memory management.} We group these baselines by model scale, including models below 10B and around 30B parameters. These agents follow a standard ReAct-style framework and run without explicit memory management under their reported context limits.

(2) \textbf{Existing memory management approaches.} We compare with summary-based agents that compress context after reaching a context threshold, as well as folding-based agents, including SWE-Compressor~\cite{DBLP:journals/corr/abs-2512-22087} and Context Folding~\cite{DBLP:journals/corr/abs-2510-11967}.

(3) \textbf{Our method variants.} We report several configurations. Workflow-only applies the proposed memory-management workflow at inference time to the untuned base model, with compression decisions and arguments provided by the proprietary model. SFT and RL variants are trained models evaluated under different step limits. In these settings, no workflow or proprietary-model assistance is used during inference.

\subsection{Experiment Details}

We use Qwen3-Coder-30B-A3B-Instruct~\cite{DBLP:journals/corr/abs-2505-09388} and Qwen3-4B-Instruct-2507~\cite{DBLP:journals/corr/abs-2505-09388} as base models, adopt OpenHands~\cite{OpenHands2025} as the base scaffold, and set the context limit to 32,768 tokens throughout data collection, SFT, and RL. For memory-management baselines, all methods use a maximum of 500 steps unless a different step budget is explicitly indicated in parentheses. Detailed hyperparameters and framework settings are provided in Appendix~\ref{sec:appendix_training_details}.

\subsection{Main Results}

Table~\ref{tab:model_performance} shows that \textsc{SWE-MeM} consistently outperforms baseline methods at both the 4B and 30B scales while using a compact 32K context limit.

In the 4B setting, \textsc{SWE-MeM} achieves a 41.6\% resolve rate with SFT and further improves to 43.4\% with RL, outperforming all ReAct-agent baselines in this scale range. In the 30B setting, the Workflow-only configuration already achieves a 58.4\% resolve rate without model training, substantially outperforming the ReAct baseline using Qwen3-Coder-30B-A3B. This result shows that effective memory management can substantially improve long-horizon software engineering agents by preserving task-relevant information and reducing distraction from redundant context.

After training, \textsc{SWE-MeM} internalizes this capability and no longer relies on an external LLMs at inference time. It reaches 58.8\% after SFT and further improves to 60.2\% after RL. Compared with existing memory-management baselines, \textsc{SWE-MeM} also shows consistent gains. For example, on their respective base models, Context Folding improves Seed-OSS-36B from 55.2\% to 58.0\%, whereas \textsc{SWE-MeM} improves Qwen3-Coder-30B-A3B from 51.6\% to 58.8\% after SFT and 60.2\% after RL. These results suggest that \textsc{SWE-MeM} provides stronger performance gains than prior memory management approaches.




\FloatBarrier

\section{Further Analysis}\label{sec:discussion}

\subsection{Efficiency Analysis}\label{sec:efficiency}

Table~\ref{tab:model_efficiency} compares the efficiency of \textsc{SWE-MeM} with existing memory-management baselines. \textsc{SWE-MeM} achieves the best task performance while maintaining the lowest reported token usage among these methods. It also uses fewer tokens than the corresponding base ReAct agent, whereas most competing memory-management baselines increase total token usage. In terms of interaction steps, the increase introduced by \textsc{SWE-MeM} remains much smaller than that of Context Folding, suggesting that selective compression can reduce redundant context without inducing excessive re-exploration.

\subsection{Ablation Study}

\begin{figure}[t]
    \centering
    \begin{minipage}[t]{0.52\linewidth}
        \centering
        \includegraphics[width=\linewidth,trim={0 440 0 0},clip]{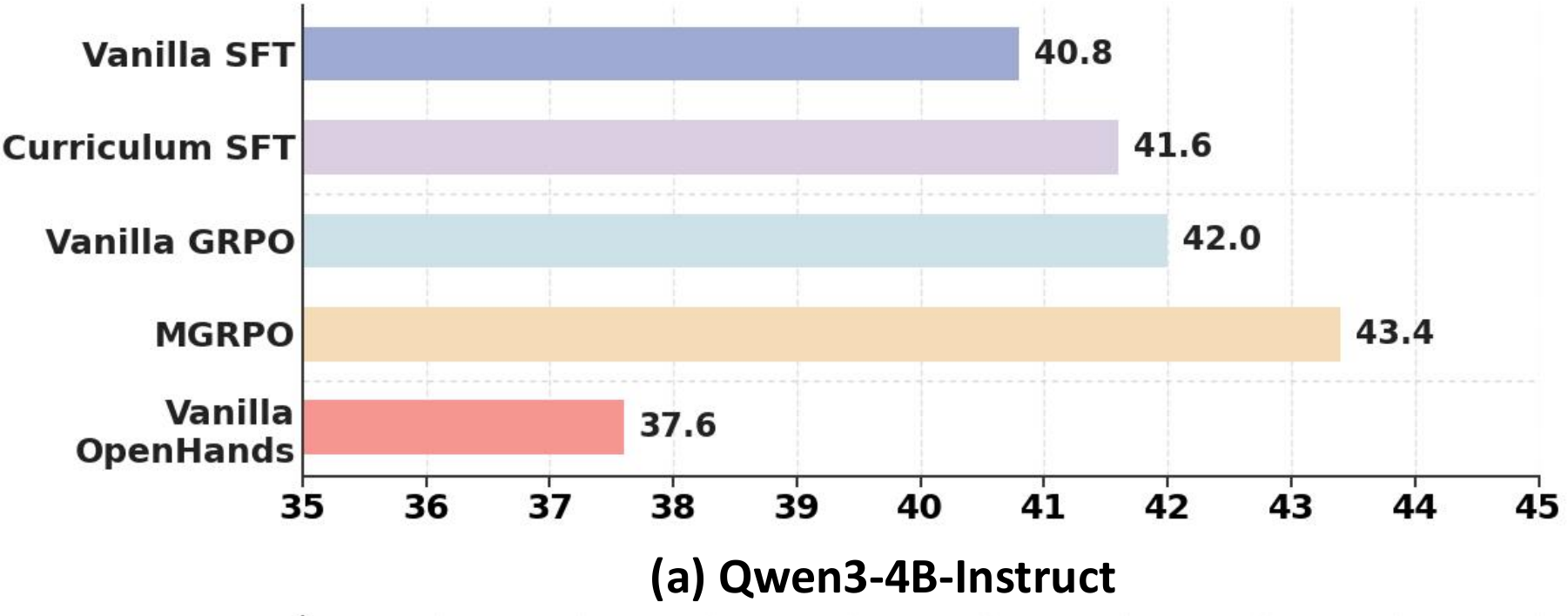}
    \end{minipage}\hfill
    \begin{minipage}[t]{0.48\linewidth}
        \centering
        \includegraphics[width=\linewidth,trim={0 0 0 355},clip]{figures/ablation.pdf}
    \end{minipage}
    \caption{Ablation study.}
    \label{fig:ablation}
\end{figure}

\textbf{Analysis of agent design.}
We ablate two key design choices by replacing proactive invocation with threshold-based invocation and replacing selective span selection with full-trajectory summarization. We conduct this analysis only on the 30B model, as the original 4B model is too weak to provide a stable comparison. As shown in Figure~\ref{fig:ablation}, replacing these components reduces the resolve rate to 57.4\% and 57.6\%, respectively. These results show that both proactive triggering and selective compression are important for final performance.

\textbf{Analysis of curriculum SFT.}
As shown in Figure~\ref{fig:ablation}, curriculum SFT consistently outperforms vanilla SFT, which uses only the first-stage training. Specifically, performance increases from 40.8\% to 41.6\% for the 4B model. In addition, after the second-stage training, the proportion of proactive tool invocations increases from 2.3\% to 16.1\% for the 4B model and from 4.1\% to 14.7\% for the 30B model. These results suggest that the curriculum design helps the model acquire more reliable proactive memory-management behavior.

\textbf{Analysis of memory-aware GRPO.}
Figure~\ref{fig:ablation} also shows that Memory-aware GRPO consistently outperforms vanilla GRPO on both model scales, indicating that loss aggregation and step-level credit assignment are important for long-horizon RL training.

\textbf{Analysis of memory management tool.}
We further evaluate the role of the memory tool by running the trained model with the vanilla OpenHands scaffold and a much larger 256K context window. After removing the memory tool, the model's performance drops substantially despite the larger context budget. For example, on the 4B model, the resolve rate decreases from 43.4\% to 37.6\%. This suggests that memory management does more than prevent context overflow: it provides a cleaner working context and helps the model use task-relevant information more effectively.

\subsection{Performance on SWE-Bench Multilingual and SWE-Bench Pro}

We further evaluate \textsc{SWE-MeM} on two challenging benchmarks: SWE-Bench Multilingual, which covers multiple programming languages, and SWE-Bench Pro, which focuses on longer-horizon tasks. As shown in Table~\ref{tab:pro}, \textsc{SWE-MeM} consistently improves over the corresponding base models on both benchmarks despite using a smaller 32K context limit instead of the 256K context used by the base models. On SWE-Bench Multilingual, \textsc{SWE-MeM} improves the resolve rate from 7.3\% to 19.0\% with the 4B model and from 35.3\% to 40.7\% with the 30B model. On SWE-Bench Pro, it improves the resolve rate from 2.6\% to 15.2\% with the 4B model and from 28.9\% to 31.7\% with the 30B model. Overall, these results suggest that the learned memory-management behavior transfers across programming languages and provides clear gains on longer-horizon software engineering tasks.

\begin{table}[t]
\centering
\resizebox{0.5\linewidth}{!}{%
\begin{tabular}{lcccc}
\toprule
\textbf{Model} & \textbf{Length} & \begin{tabular}[c]{@{}c@{}}\textbf{SWE-bench}\\\textbf{Multilingual}\end{tabular} & \begin{tabular}[c]{@{}c@{}}\textbf{SWE-bench}\\\textbf{Pro}\end{tabular} \\
\midrule
Qwen3-4B-Instruct & 256k & 7.3 & 2.6 \\
+ \textsc{SWE-MeM} SFT+RL & 32k & 19.0 & 15.2 \\
\midrule
Qwen3-Coder-30B-A3B & 256k & 35.3 & 28.9 \\
+ \textsc{SWE-MeM} SFT+RL & 32k & 40.7 & 31.7 \\
\bottomrule
\end{tabular}
}
\caption{Experimental results on SWE-bench Multilingual and SWE-bench Pro.}
\label{tab:pro}
\end{table}

\FloatBarrier
\section{Related Work}\label{sec:related}

\subsection{Software Engineering Agents}
Software engineering agents build on a progression from atomic code-intelligence capabilities~\cite{CodeBERT,GraphCodeBERT,GaoWGWZL23,Gao0GL25} to autonomous repository-level task solving~\cite{SWEAgent2024,OpenHands2025,Agentless2024,DBLP:journals/corr/abs-2507-23348}. Early capability-oriented work studied code generation~\cite{DBLP:journals/corr/abs-2107-03374,DBLP:journals/corr/abs-2203-07814,CodeT5,abs-2510-17130,DBLP:journals/corr/abs-2601-00376,DBLP:journals/corr/abs-2410-01215}, program repair~\cite{XiaWZ23,PengGGHL24}, code summarization~\cite{DBLP:journals/corr/abs-2601-05485} ,code translation~\cite{RoziereLCL20,LuoJGGFLXL25}, and test generation~\cite{abs-2501-01329,chen2024chatunitest}. Later work explored agent scaffolds for repository-level problem-solving, including SWE-Agent, OpenHands, and Agentless~\cite{SWEAgent2024,OpenHands2025,Agentless2024}. Recent progress has further broadened the ecosystem through richer environments and benchmarks~\cite{SWEGym2025,SWERebench2025,SWEFactory2025,DBLP:journals/corr/abs-2603-20691,DBLP:journals/corr/abs-2602-02361,DBLP:journals/corr/abs-2602-03419}, synthetic data generation~\cite{SWEMirror2025,BugPilot2025}, training or post-training recipes~\cite{DBLP:journals/corr/abs-2602-03411,SWERL2025,SWESwiss2025,DeepSWE2025}, reward or verifier modeling~\cite{DBLP:journals/corr/abs-2512-21919}. Our work complements this line by focusing on how agents should learn to manage context during long-horizon tasks.

\subsection{Memory Management}
Managing context in extended interactions is critical for agents operating long-horizon tasks across hundreds of rounds. Early context compression approaches include token-level pruning methods like LLMLingua~\cite{LLMLingua2023} and code-specific optimization techniques such as DietCode~\cite{DietCode2022}. More recent work has rapidly expanded this direction with context folding and routing, adaptive pruning, trajectory reduction, and trainable context-management policies for web and coding agents~\cite{DBLP:journals/corr/abs-2508-05988,DBLP:journals/corr/abs-2603-27490,AgentFold2025,DBLP:journals/corr/abs-2601-16746,DBLP:journals/corr/abs-2601-18285,SUPO2025,ContextPilot2026,DBLP:journals/corr/abs-2601-05110,DBLP:conf/kbse/ShiQZSG25}. For example, AgentDiet reduces wasteful historical context through trajectory reduction~\cite{DBLP:journals/corr/abs-2509-23586}, Context Folding summarizes completed subtasks and folds the associated action records into compact subtask-level summaries~\cite{DBLP:journals/corr/abs-2510-11967}, while SWE-Compressor elevates context maintenance to a callable tool integrated into the agent's decision-making process~\cite{DBLP:journals/corr/abs-2512-22087}, enabling SWE agents to compress context at appropriate milestones rather than relying on passive or externally-triggered compression strategies. Compared with methods that rely on fixed compression rules or externally defined scopes, \textsc{SWE-MeM} trains the agent to decide when to compress, what span to compress, and how to rewrite it under the current trajectory state.

\FloatBarrier


\section{Conclusion}\label{sec:conclusion}

We presented \textsc{SWE-MeM}, a framework for training long-horizon SWE agents with proactive and on-demand memory management. By combining a flexible memory tool, trajectory synthesis with curriculum fine-tuning, and Memory-aware GRPO, \textsc{SWE-MeM} improves SWE-Bench Verified performance, reaching 43.4\% and 60.2\% resolve rate in our 4B and 30B settings, respectively, and outperforming existing memory management baselines in both performance and efficiency.
 

\FloatBarrier
\bibliographystyle{plainnat}
\bibliography{references}

\clearpage
\appendix
\section*{Appendix}
\addcontentsline{toc}{section}{Appendix}

\section{Method Details}\label{sec:method_detail}

\subsection{Detailed Trajectory Synthesis Procedure}
Algorithm~\ref{alg:trajectory-synthesis} describes the details of memory-management trajectories construction process. 

\textbf{Step 1: rollout initialization.} As shown lines 1--3 of Algorithm~\ref{alg:trajectory-synthesis}, we initialize the conversation with the task prompt, the context budget, the base model, and the Proprietary model. The base model is used for generating ordinary agent actions, while the Proprietary model is only used when we need synthetic memory-management supervision. This separation help us study to what extent training that focuses solely on memory management can improve the SWE agent's capabilities.

\textbf{Step 2: budget-aware monitoring.} At each iteration, lines 4--7 compute the remaining token budget and initialize the compression state. Lines 8--12 then decide whether the current trajectory should trigger compression. The budget-pressure branch in lines 8--10 uses the remaining-budget heuristic $p_t$ to increase compression probability as the context window becomes saturated. The proactive branch in lines 11--12 corresponds to the context-monitoring judge described by the \hyperref[prompt:context-monitor]{Context Monitor Prompt}. In practice, we do not query the Proprietary model at every step because doing so would be unnecessarily expensive and most states are not suitable for proactive compression. Moreover, patterns such as \textit{subtask end} do not always yield a useful compression opportunity, so the goal of SFT cold start is only to expose the model to basic proactive-memory behavior rather than to enforce compression at every possible timing. Concretely, after each step we query the monitor with only 10\% probability; in the remaining 90\% steps, proactive compression is skipped entirely. We further skip the monitor if any compression action has occurred within the most recent five steps, since the required conditions are unlikely to be satisfied again at such short intervals.

\textbf{Step 3: generating the compression action.} Once compression is triggered, lines 9--12 call the Proprietary model to synthesize the actual memory operation. The generated action follows the \hyperref[prompt:compression-generation]{Compression Generation Prompt}, which asks the model to first reflect on current progress, then select a valid compression span, and finally produce the compressed content together with a concrete remaining-work plan. The prompt also constrains the selected span to avoid illegal overlaps with previously compressed ranges and encourages the model to preserve information that will still be needed in later rounds.

\textbf{Step 4: trajectory update and continued rollout.} After the compression action is synthesized, line 14 updates the context according to Eq.~(2): the selected source span is replaced by compressed content and the remaining-work summary is appended to the end of the trajectory. If no compression is triggered, lines 15--16 simply let the base model continue interacting with the environment.

\begin{algorithm}[h]
\small
\caption{Trajectory Synthesis with Dynamic Memory Management}
\label{alg:trajectory-synthesis}
\begin{algorithmic}[1]
\STATE \textbf{Input:} Task prompt $p$, max iterations $T$, context budget $B$, local trajectory model $\mathcal{M}_{\text{local}}$, Proprietary model $\mathcal{M}_{\text{prop}}$
\STATE \textbf{Output:} Trajectory with memory operations $\tau = \{(c_t, a_t, o_t)\}_{t=1}^{T}$
\STATE Initialize conversation history $C \leftarrow [\text{system}, \text{user}(p)]$
\WHILE{$t < T$ and task not finished}
    \STATE $t \leftarrow t + 1$, compute remaining budget $b_t \leftarrow B - \text{TokenCount}(C)$
    \STATE $\text{NeedCompression} \leftarrow \texttt{False}$
    \STATE $(r_{\text{start}}, r_{\text{end}}, c, g) \leftarrow (\varnothing, \varnothing, \varnothing, \varnothing)$
    \STATE \textcolor{blue}{// \textbf{Category (1): Context Budget Pressure}}
    \STATE $p_t \leftarrow \max\!\left(0,\min\!\left(1,\frac{0.20B-b_t}{0.15B}\right)\right)$
    \IF{$u \sim \mathrm{Uniform}(0,1) < p_t$}
        \STATE $(r_{\text{start}}, r_{\text{end}}, c, g) \leftarrow \mathcal{M}_{\text{prop}}.\text{BuildCompression}(C)$
        \STATE $\text{NeedCompression} \leftarrow \texttt{True}$
    \ELSIF{$\text{NeedProactiveCompression}(C)$}
        \STATE \textcolor{blue}{// \textbf{Category (2): Proactive Memory Management}}
        \STATE $(r_{\text{start}}, r_{\text{end}}, c, g) \leftarrow \mathcal{M}_{\text{prop}}.\text{BuildProactiveCompression}(C)$
        \STATE $\text{NeedCompression} \leftarrow \texttt{True}$
    \ENDIF
    \IF{$\text{NeedCompression}$}
        \STATE Update $C$ following Eq.~(2)
    \ELSE
        \STATE $a_t \leftarrow \mathcal{M}_{\text{local}}(C)$ \COMMENT{Generate agent action}
        \STATE Execute $a_t$, observe $o_t$, update $C$ and record $\tau$
    \ENDIF
\ENDWHILE
\STATE \textbf{return} $\tau$
\end{algorithmic}
\normalsize
\end{algorithm}

\subsection{Detailed Description of Trajectory Splitting Process}

Suppose memory-management actions are triggered at turns $\{t_1,\ldots,t_{K-1}\}$, with $0=t_0<t_1<\cdots<t_{K-1}<t_K=T$. For the $(k\!-\!1)$-th memory event, let $(s_k,e_k,c_k,f_k)$ denote the selected compression span and generated summary contents, where only the source range $[s_k,e_k]$ is replaced while all earlier and later context outside this span is retained. Following Eq.~\eqref{equ:compression}, the compressed context after this event can be written directly as
\begin{align}
C'_{t_k} = \big\langle
&p, \{a_1, o_1\}, \ldots, \{a_{s_k-1}, o_{s_k-1}\}, \nonumber \\
&\{c_k\}, \{a_{e_k+1}, o_{e_k+1}\}, \ldots, \nonumber \\
&\{a_{t_k-1}, o_{t_k-1}\}, \{f_k\}
\big\rangle.
\end{align}
Thus, the post-compression context is a mixed trajectory consisting of retained original rounds, one compressed span, and the appended remaining-work item, rather than a single summary of the whole prefix.

We then split the full rollout at compression boundaries so that each segment starts from exactly the compressed context that would be available at deployment time:
\begin{align}
\tau^{(k)} = \big\langle
&C'_{t_{k-1}}, \{a_{t_{k-1}+1}, o_{t_{k-1}+1}\}, \ldots, \nonumber \\
&\{a_{t_k}, o_{t_k}\}
\big\rangle,
\quad k=1,\ldots,K.
\end{align}
Equivalently, the full rollout is decomposed as $\tau \xrightarrow{\mathcal{M}} \{\tau^{(1)}, \tau^{(2)}, \ldots, \tau^{(K)}\}$. Here $C'_{t_{k-1}}$ already preserves all uncompressed context outside the selected source span of the previous memory event, so each sub-trajectory $\tau^{(k)}$ is conditioned on the same partially compressed information state that the agent would actually observe online. This keeps the consistency of model training and inference state.

\subsection{Prompts}\label{sec:prompt}

\paragraph{Context Monitor Prompt}\label{prompt:context-monitor} Prompt used by the context monitor to decide whether compression should be triggered.
\begin{PromptBox}
You will be provided with a complete trajectory of an AI agent resolving a software bug. Your task is to determine whether compression is needed and identify the compression type.

**Input Trajectory:**
<trajectory>
{text_conversation}
</trajectory>

Please evaluate whether the trajectory meets ANY of the following conditions for compression:

**Condition 1: Subtask Completion with Redundant Details**
- The last action marks the **completion** of a whole phase  rather than an intermediate step or the **start** of a new phase. The next logical action would be to proceed to a subsequent phase
- All errors encountered during this phase have been resolved and validated
- **CRITICAL EXCLUSION**: If the last step shows the agent **entering** a new phase without completing substantive work, this does NOT qualify as subtask completion
- Do NOT choose `subtask_end` if the last 1-3 rounds show unresolved blocker evidence, active debugging of bugs, or unvalidated regressions
- **CRITICAL: Low overall information density**:
  - The completed subtask is lengthy but **overall information density is low**
  - **Only key conclusions or final states** (e.g., "bug found in file X", "all tests pass") will be referenced in subsequent steps
  - **Intermediate details are NOT needed for future actions**: Future steps only need the "what was accomplished" rather than the "how it was accomplished"

**Condition 2: Low Information Density with No Future Utility**
Evaluate recent actions (only consider last 1-5 rounds, don't consider earlier rounds) for lengthy outputs with low information value that will NOT be needed in subsequent steps:
- **Successful file modification**: → HIGH density (don't compress)
- **Early exploration phase**: Examining files/directories → relatively HIGH density (they are likely to be referenced in future steps, so they should be carefully decided whether to compress)
- **Untested test cases that reproduce the bug or examine the fix** → HIGH density (don't compress)
- **Script execution or test results**:
  - Execution succeeded with verbose output unlikely to be referenced again → LOW density
  - Execution failed with useful error messages needed for debugging → HIGH density (don't compress)
  - Execution failed with configuration/usage errors already resolved → LOW density
- **Lengthy failed execution with simple failure reason** that won't inform future decisions → LOW density
- **Long thinking process** only the conclusion is useful → LOW density
- **Key criterion**: The content should be compressible because its information density is low and future actions won't need to refer to this step's detailed content
- **REQUIRED: Standalone/isolated content**: Recent rounds must contain self-contained, one-off tasks completely unrelated to subsequent workflow and future tasks. If the content is part of an ongoing workflow or may be referenced in future steps, do NOT compress regardless of information density.
- **CRITICAL: Consecutive low-density requirement**: Only compress when the last 3-5 rounds (counting backwards from the most recent round) are ALL low density. If any round within this range is high density or potentially useful for future steps, do NOT compress at all. Prefer compressing 3-5 consecutive low-density rounds together. Don't just compress for only one or two rounds unless they are too long.

**Condition 3: Focus Degradations Requiring Agent Alert**
**ONLY evaluate this condition AFTER a patch has been generated.** Before patch generation, ignore this condition entirely.
Use `critical_issue` conservatively:
- Do NOT use `critical_issue` only because a problem just appeared; the agent may fix it in the next few steps.
- Do NOT use `critical_issue` only because tests have not been run yet; the agent may still test later.
- Do NOT use `critical_issue` based on generic risk wording without concrete, recent evidence.
- Use `critical_issue` when the same core issue remains unresolved across multiple recent attempts, or when repeated attempts still fail to fix it.
- Use `critical_issue` when the patch itself shows clear correctness risk (e.g., wrong edit location, broken boundary conditions, contradictory logic, or likely regression) and the agent is not correcting it.
- Use `critical_issue` when the agent is moving to final review / summarization / submission while key risks are still unresolved.
- Do NOT repeat the same `critical_issue` in consecutive decisions unless new evidence appears.
- Require concrete evidence: include specific failing test/error/traceback or explicit wrong behavior in recent rounds.
- Require persistence: the core issue should remain unresolved across at least two recent attempts (or one attempt plus clear prior failed attempt evidence).
- If the same issue was already flagged and there is no new evidence in recent rounds, return `should_compress: false`.

Compress with special alert if ANY of the following serious problems exist:

**1. Patch Quality Issues with High Correctness Risk**
The generated patch itself contains clear flaws that pose significant correctness risks:
- Wrong edit location or scope (modifying unrelated code sections)
- Broken boundary conditions or edge case handling
- Contradictory logic or likely regressions
- **AND** the agent shows no awareness of these critical flaws or is not attempting to correct them

**2. Unresolved Core Issues After Multiple Attempts**
The same fundamental problem persists across multiple iterations:
- The agent has made repeated attempts but the core issue remains unresolved
- Validation or tests continue to fail for the same underlying reason
- The problem is not merely "not fixed yet" but shows a pattern of ineffective approaches

**3. Premature Finalization with Unaddressed Risks**
The agent is moving toward completion while critical risks remain:
- Entering final review, summarization, or submission phase
- While the patch has unresolved correctness concerns that were identified in previous steps, or tests may pass (or haven't been run) but there are untested edge cases, likely impact on related code paths, or the test coverage is clearly inadequate for the scope of changes


**IMPORTANT**: If Condition 3 applies, you should still compress and use type `"critical_issue"` to alert the agent about these problems.

**Output Format:**

<analysis>
[Your detailed analysis evaluating all conditions and determining which (if any) applies]
[If compression is needed, explain which condition triggered it]
[DO NOT specify the exact round numbers - this will be determined in the next step based on detailed reflection]
</analysis>

<decision>
[Your decision in JSON format]

**Important**: When compression is needed, write the reason field in first-person perspective describing why you want to compress the content, based on the compression type. Describe naturally what happened without following a fixed template.

If compression is needed:
{{
    "should_compress": true,
    "type": "subtask_end" | "low_density" | "critical_issue",
    "reason": "[First-person description of what you just completed and why it can be compressed]"
}}

If no compression is needed:
{{
    "should_compress": false,
    "type": "none",
    "reason": "[Brief explanation why compression is not needed]"
}}
</decision>
\end{PromptBox}

\paragraph{Compression Generation Prompt}\label{prompt:compression-generation} Prompt used to generate the compressed content and remaining-work plan.
\begin{PromptBox}
You are an AI agent working on resolving a software bug. You need to reflect on your debugging progress and compress portions of your conversation history to maintain context efficiency while preserving critical information.

{compression_context}

## Context

Here is your complete debugging trajectory so far:

<trajectory>
{text_conversation}
</trajectory>

## Your Tasks

Complete the following tasks in order:

### Task 1: Comprehensive Reflection and Compression Planning

Write a natural, flowing reflection about your debugging process—think of it as your internal thoughts. This should be a cohesive narrative (not bullet points or a list of answers) that weaves together:

**Part 1: Current Status Review**

- Briefly summarize what you have accomplished in the current trajectory, referencing specific round numbers for key milestones. The summarization should be very concise and focus on the key achievements. Do not describe too much details about each part.
- Taking a critical perspective, carefully review the current solution to identify whether it contains any of the following problems: incorrect issue understanding, incorrected issue location (Truly identify the root problem, or is it just a superficial error influenced by the current test case? Is there any verification?), previously failing tests that were left unresolved before moving forward, tests that failed to run and were simply skipped or abandoned, fixes that do not represent the minimal necessary changes, patches that modify aspects beyond the issue description (such as inconsistent boundary handling that may introduce regressions), or after modification not write more tests to verify various edge cases (such as empty or none input). Make your own independent judgment about each problem type—this is not a reading comprehension task where you summarize what have been found in the trajectory; you need to actively detect these issues based on your own analysis.

**Part 2: Future Task Planning and Information Requirements**

- What are your pending tasks and what specific actions are you planning to take in the next few steps? (Note that you cannot engage in any communication with users and should only focus on resolving this issue without any interaction with users.)
- **Information Requirements Analysis**: Before selecting compression ranges, explicitly think through: If you were to hand off these pending tasks to another person who can only see the compressed trajectory, what specific information would they absolutely need to complete each action successfully? For example:
  - If writing a patch: They need the original file content, the exact location of the bug, and the root cause explanation
  - If running tests: They need to understand what each test case is validating and why
  - If implementing a fix: They need the context of how the code is used and what edge cases to consider
  Naturally articulate these information requirements in your reflection before moving to compression selection.

**Part 3: Compression Range Selection**

Based on the information requirements for your future actions, identify which range of rounds should be compressed. Your selection should balance the positioning of task-relevant content with compression efficiency, prioritizing ranges with a lower proportion of information that will be needed for pending tasks. However, this doesn't mean you can't include such content—if your selected range contains information that will be needed later, that's acceptable, as you can choose to preserve it verbatim. Your selected range must satisfy the following Mandatory Requirements and carefully consider the Selection Priorities:

**Mandatory Requirements:**
- You cannot select the entire trajectory range from start to end
- First list existing compressed ranges only from previous user compression messages (content in <compressed_content>, specifically {existing_ranges_text}). Do NOT treat merged round titles inside compressed text (e.g., "Rounds 4-7 (Merged)") as existing compressed ranges. The new range must be either disjoint, or a full combination that fully covers any existing compressed ranges. Do not choose identical ranges or partial overlaps.
  - Allowed: existing [11-20], [21-30] -> new [11-30] (combine two adjacent ranges), or existing [11-20], [21-30] -> new [21-35] (combine compressed content with uncompressed content)
  - Not allowed: existing [11-20] -> new [15-25] (overlap), or new [11-20] (identical)
- The issue description cannot be compressed, so the starting round number must be at least 1
- The compression range should cover approximately 30% of the total trajectory token length. Avoid selecting too few rounds at once (generally no fewer than 5 rounds)

**Selection Priorities:**
- Try not to compress the last few rounds that are still in progress
- Prioritize compressing earlier rounds over later ones when possible
- Prioritize selecting ranges containing:
  1. Lengthy code execution outputs with minimal relevance to subsequent steps
  2. Large file openings or directory listings where only limited information is utilized later
  3. Consecutive rounds completing a phase where intermediate results have no bearing on subsequent processes
  4. Steps from many rounds earlier that have little impact on the most recent round
  5. Multiple adjacent existing compressed segments that can be combined

**Part 4: Content Importance Assessment**

For your selected range, assess the importance level of each piece of content:

- **Need preserve verbatim**:  Content that might be referenced by future steps, such as test cases revealing the bug, error traces showing root cause, key code snippets related to the bug, the files you modified
- **Summarize while retaining details**: Content with limited ongoing relevance, such as long grep or list results, repeated trials, errors that are ultimately resolved and methods that you tried but didn't work
- **Can compress heavily with key findings/lessons**: Content unlikely to be referenced again, such as long passed logs, script execution method, or useless execution outputs

In general, you should verbatim preserve about 30% of the content in your selected range. Make sure all information relevant to future steps is preserved.

Naturally integrate these assessments into your reflection, explaining which content needs to be preserved, and which can be summarized or heavily compressed.

At the end of your reflection, naturally indicate which rounds you will compress (you don't need to use a fixed format—just make it clear which range you've selected).

### Task 2: Generate Compressed Content

Compress your selected range following these guidelines:

**1. Structure for each round:**

<example>
**Round X** or **Rounds X-Y (Merged)** (Phase 1 optional) — [Brief description]

- **Action:**
  [What was done and why—focus on purpose and outcome]

- **Results/Findings & Relevance:**
  [Important discoveries or results (either preserve verbatim or summarize), with brief note on why this information is/isn't still needed.]
</example>

- You can merge multiple adjacent rounds together to describe their content. If the range to be compressed includes content that has already been compressed, you can further expand the scope of the existing merges and appropriately reduce the description of unimportant details.
- **CRITICAL: Phase annotations is optional, but must strictly follow the Original Conversation's phase declarations:** When rounds are annotated with phase numbers (e.g., "**Rounds 1-5** (Phase 1)"), these phase labels MUST come from explicit phase declarations in the Original Conversation. If the original text does not explicitly mention phases, do NOT add phase annotations. Phase numbers should logically increment (e.g., Phase 1, Phase 2, Phase 3) and should not appear out of order.


**2. Apply compression based on importance level:**
- **Need preserve verbatim**: Preserve verbatim with full context
- **Summarize while retaining details**: Summarize results and keep key details that are relevant to future steps
- **Can compress heavily with key findings**: Key results/findings or merge with adjacent rounds

**3. Content-specific handling:**
- **Code files**: Keep essential and problem-related sections, omit boilerplate code or code comments containing lengthy but unnecessary explanations
- **Test results**: One sentence for all passing tests; for failures that are still relevant, keep key error message
- **Error tracebacks**: Keep entry point and actual error line, omit system library paths and framework internal calls
- **Repetitive Content like list or grep results**: Summarize directory contents instead of listing every file, such as "Under `/testbed/src/docx/opc/` there are: part.py, parts/, phys_pkg.py, pkgreader.py, pkgwriter.py. Under `/testbed/src/docx/oxml/` there are: __pycache__, coreprops.py, document.py, drawing.py, exceptions.py"

**4. Always exclude:**
- Content outside your selected range
- Meta information in `<NOTE>` tags
- Explanatory text about the compression process in user message

**5. Maintain continuity:**
- No placeholders like "...rest remains the same..."
- Ensure compressed content flows naturally
- Preserve enough information so future rounds remain understandable

### Task 3: Progress Summary and Forward Planning

Based on your reflection in Task 1, extract and write a concise summary that includes:

**Analysis of Past Steps:**
- Briefly state the Phases that you have completed in one sentence
- Clearly identify if there are any mistakes, oversights, or problems in your previous actions that haven't been corrected yet
- If exists, Be specific about what went wrong and why
- If no significant issues were found, state this clearly

**Future Action Plan:**
- List the specific next steps you plan to take and where in the trajectory the information needed to complete these tasks can be found (with round numbers)
- The plan here must be precise and cannot be vague. For example, you cannot write what can optionally be done, nor can you simply say "run more tests" - you must specify which cases still need to be tested.

This summary should be clear, actionable, and distilled from your reflection—focusing on concrete problems and solutions without repeating all the details from the reflection.

## Output Format

<reflection>
[Your natural narrative reflection that includes: your accomplishments with round numbers, assessment of your process, pending tasks, compression range selection with rationale, and importance assessment of content within the selected range. End by naturally stating which rounds you will compress.]
</reflection>

<start_round_number>
[integer]
</start_round_number>

<end_round_number>
[integer]
</end_round_number>

<compressed_content>
[Your compressed content for the selected range only]
</compressed_content>

<progress_summary>
[Your concise summary of issues with past steps and future action plan, extracted from your reflection above]
</progress_summary>
\end{PromptBox}

\paragraph{Stage-2 Filtering Prompt: Compression Quality}\label{prompt:stage2-compression-quality} Rubric prompt from stage-2 filtering for judging whether a synthesized compression sample should be kept.
\begin{PromptBox}
You are judging whether a context-compression trajectory is a good training sample.

Only use the full trajectory below as evidence.

Judge these five conditions as true or false:
1. reflection_identifies_process_problem_and_future_fix_plan:
   The reflection must identify real problems, weaknesses, mistakes, risks, or gaps in the current problem-solving process itself, and it must propose how the later work should change because of that. If the reflection only says progress is fine, a phase is complete, tests passed, or that it wants to compress to save context, then this condition must be false.
2. source_selection_is_good:
   The selected compressed range is worth compressing. It should not be too short in actual content volume. A small round span is still acceptable if the source itself is long or dense, such as long logs or detailed debugging content.
3. compressed_content_preserves_source_critical_information:
   The produced compressed content preserves the source-critical information for this specific source range. Judge this relative to the actual source content: for logs, preserving the key signals and conclusions may be enough; for analysis, implementation, or verification rounds, the summary should preserve the technically important facts needed for later continuation.
4. compression_plan_matches_actual_compressed_content:
   This must be judged strictly. The pre-tool analysis, the claimed compression range, and the actual compressed content must point to the same material. If the analysis says it will compress one part but the result actually summarizes another part, or if the summary mixes in out-of-range material, or if it omits the very material it said it would compress, then this condition must be false.
5. future_plan_is_specific_and_actionable:
   The future plan in the final parameter/progress-summary style content must be concrete and actionable, not generic phase language. It should name specific next checks or actions, such as running tests for a particular file/function/bug scenario, checking a specific behavior, or verifying a specific risk. Generic statements like 'continue verification', 'move to the next phase', or 'keep working' are not enough.

Return strict JSON only with this schema:
{
  "reflection_identifies_process_problem_and_future_fix_plan": true,
  "source_selection_is_good": true,
  "compressed_content_preserves_source_critical_information": true,
  "compression_plan_matches_actual_compressed_content": true,
  "future_plan_is_specific_and_actionable": true,
  "failed_conditions": ["..."],
  "reason": "brief explanation"
}

If any condition is not satisfied, set it to false and name it in failed_conditions.
Do not output scores.

Full trajectory:
{sample.full_trajectory_rendered or '[EMPTY]'}
\end{PromptBox}

\paragraph{Stage-2 Filtering Prompt: Split Quality}\label{prompt:stage2-split-quality} Rubric prompt from stage-2 filtering for judging post-compression trajectory quality.
\begin{PromptBox}
You are judging whether this trajectory is a good training sample.

Judge these conditions as true or false:
1. no_bad_assistant_content:
   Assistant messages should NOT be bad content such as obvious repeated output, repetitive low-value text, malformed garbage, bizarre or corrupted characters, broken formatting artifacts, or other clearly abnormal content that should not be learned.
2. post_reflection_actions_follow_future_plan:
   After the reflection/compression user message, the later assistant actions should substantially follow the future plan implied by that reflection and compressed summary. It is okay if the agent adapts slightly, but it should not ignore the future plan and jump to unrelated work.
3. no_unnecessary_repeated_actions_after_compression:
   After the reflection/compression user message, there should NOT be clearly unnecessary repeated actions caused by forgetting compressed context. For example, if the earlier trajectory already established some file contents, logs, test results, or conclusions, the later assistant should not redundantly redo the same file viewing, same test runs, or same checks without a clear new reason.

Return strict JSON only with this schema:
{
  "trajectory_type": "with_compressed_user_message",
  "no_bad_unmasked_assistant_content": true,
  "post_reflection_actions_follow_future_plan": true,
  "no_unnecessary_repeated_actions_after_compression": true,
  "failed_conditions": ["..."],
  "reason": "brief explanation"
}

If any condition is not satisfied, set it to false and include it in failed_conditions.
Do not output scores.

Full trajectory:
{sample.rendered_trajectory or '[EMPTY]'}
\end{PromptBox}

\section{Experiment Details}

\subsection{Dataset Details}\label{sec:appendix_dataset_details}
\textbf{Training Data.} We construct the SFT and RL data from SWE-ReBench \cite{SWERebench2025} and SWE-Gym \cite{SWEGym2025}. For SFT data synthesis, we use Qwen3-Coder-30B-A3B-Instruct as the base model and GPT-5.1 as the proprietary synthesis model, perform three rollouts per problem, and apply the filtering pipeline described in Section~\ref{subsec:fitering}. For fully correct problems, we keep only one trajectory to avoid overweighting overly simple instances. The resulting SFT data statistics are summarized in Table~\ref{tab:data_stats}.

For RL, we further construct a filtered non-trivial subset from the same two sources. We first run ground-truth evaluation and remove instances whose reference patches fail the associated test cases. We then apply best-of-n verification to retain instances with intermediate pass rates so that the RL training set avoids both trivially easy and completely intractable samples. This procedure yields 2,194 training instances.

\textbf{Evaluation Data.} We evaluate on SWE-Bench-Verified \cite{SweBench2024}. We also evaluate on SWE-bench Multilingual, which extends SWE-bench-style evaluation to repositories in multiple programming languages, and SWE-bench Pro, which emphasizes more difficult long-horizon issue-resolution tasks; for both benchmarks, we use the full benchmark sets. In the evaluation environment, we remove future commits to avoid leakage from unintended repository history. We report issue resolve rate as the main evaluation metric.

\subsection{Training Details}\label{sec:appendix_training_details}
We use Qwen3-Coder-30B-A3B-Instruct and Qwen3-4B-Instruct-2507 as the base models for all experiments, adopt OpenHands as the agent scaffold, and set the context limit to 32,768 tokens throughout data collection, SFT, and RL.

\textbf{Supervised Fine-Tuning.} We train the SFT models with the VeOmni framework~\cite{VeOmni2025} using a global batch size of 128, a learning rate of \(2\times10^{-5}\), a cosine learning-rate schedule, and 3 training epochs. We then run a second-stage curriculum tuning pass on the filtered proactive-memory subset with learning rate \(2\times10^{-6}\) for 1 epoch. For Qwen3-4B, we directly reuse the training data collected by Qwen3-Coder-30B-A3B-Instruct.

\textbf{Reinforcement Learning.} We train the RL models with the Verl framework~\cite{HybridFlow2025} using a batch size of 64, 8 rollout samples per instance, learning rate \(2\times10^{-6}\), and \(\tau\)=0.05, matching the low-temperature setting used during trajectory synthesis. We follow the DAPO clipping setting~\cite{DAPO2025} with clip low 0.2 and clip high 0.27, and set the rollout temperature to 1.0. We also enable R3 Rollout Routing Replay to mitigate training-inference inconsistency during RL, and apply it only in the 30B RL training setting. To control training cost, we set the max interaction steps as 150 during RL. Error samples that exceed this maximum step limit are filtered and masked out during training. We do not use entropy regularization or KL-divergence penalties in the RL objective.

\begin{table}[t]
\centering
\scalebox{0.8}{
\begin{tabular}{lc}
\toprule
\textbf{Statistic} &  \\
\midrule
\textit{Data size} & \\
\quad Sample number & 8,728 \\
\quad Unique problem number & 5,867 \\
\quad Total sub-trajectory number & 27,088 \\
\quad Avg sub-trajectories per sample & 3.10 \\
\midrule
\textit{Compression} & \\
\quad Avg tokens before & 9,657.93 \\
\quad Avg tokens after & 1,496.52 \\
\quad Avg ratio (\%) & 15.50 \\
\bottomrule
\end{tabular}
}
\caption{Statistics for the SFT dataset.}
\label{tab:data_stats}
\end{table}

\section{Case Study}
In this section, we present some cases on the memory management tool. We first show several examples of real memory management invocations, and then illustrate several complete trajectories that contain memory management actions.

\subsection{Memory Management Action Examples}
We show four real memory management actions. Each box includes the short assistant rationale immediately before compression and the complete tool call. These four action examples do not overlap with the trajectory cases in the next subsection.

\paragraph{Example 1: Context-Budget Pressure. Instance: \texttt{django\_\_django-11551-0}. Step: 38.}
\begin{tcolorbox}[
  breakable,
  enhanced,
  width=\linewidth,
  colback=gray!4,
  colframe=gray!60,
  boxrule=0.5pt,
  arc=1.2mm,
  left=0.6mm,
  right=0.6mm,
  top=0.8mm,
  bottom=0.8mm,
]
\begin{lstlisting}[
  basicstyle=\tiny\ttfamily,
  breaklines=true,
  breakatwhitespace=false,
  breakautoindent=false,
  breakindent=0pt,
  columns=fullflexible,
  keepspaces=false,
  showstringspaces=false,
  upquote=true,
  tabsize=2,
  basewidth=0.46em,
  inputencoding=utf8,
  extendedchars=true,
  literate={→}{{$\rightarrow$}}1 {—}{{--}}1
]
[analysis]
So far in this trajectory I've gone from initial orientation to a concrete code change, but the fix still isn't behaving as the issue description's "correct behaviour" example promises. In round 1 I restated the problem and constraints. Rounds 2--7 (already compressed) covered repo inspection and confirming this is a Django source tree with tests and no `manage.py`. In rounds 8--14 I located `_check_list_display_item` in `django/contrib/admin/checks.py`, inspected the current implementation, and then used `git show` to compare it with the historical commit that broke the behaviour. By round 22--25 I had a small reproduction script (`simple_reproduce.py`) that demonstrated `admin.E108` being raised for a `MockPositionField`-like descriptor. In rounds 30--34 I added more debug scripts (`debug_reproduce.py`, `debug_reproduce2.py`) to understand how `hasattr`, `getattr`, and `_meta.get_field` behave for the descriptor. Finally in rounds 37--38 I reworked `_check_list_display_item` again, aligning it more closely with the "correct behaviour" snippet from the issue description.

Looking critically at the current solution, there are several issues. First, my understanding of the root cause is broadly correct--the bug stems from the 28490-fix that introduced a `hasattr(model, item)` guard that blocks `_meta.get_field(item)` for instance-only descriptors--but I haven't yet achieved the desired behaviour. The latest `_check_list_display_item` (round 38) now tries `_meta.get_field(item)` first, falls back to `getattr(model, item)` on `FieldDoesNotExist`, and only errors if that fails. However, my debug script in round 30 showed that `getattr(Thing, 'order')` raises `AttributeError: PositionField can only be accessed on model instances`, which means my fix still ends up hitting the E108 path instead of treating this as a valid callable/attribute. That indicates my logic conflates "no field at all" with "invalid list_display entry"; the correct behaviour should treat anything callable or attribute-like as valid, regardless of whether `_meta.get_field` finds a real Django field. So Phase 5 is incomplete--my fix is still wrong relative to the issue's expectations. 

Second, I haven't run any of Django's own tests yet; only my standalone scripts have been executed. That means I have no assurance about regressions in the broader admin checks or anywhere else that `_check_list_display_item` is used. There are no skipped or abandoned failing tests; I simply haven't invoked the test runner at all, which is an oversight relative to good practice. Third, my patch may be slightly less minimal than ideal because it changes the control flow around `FieldDoesNotExist` and `AttributeError` in a way that might not perfectly mirror the historical behaviour, but it is still localized to `_check_list_display_item`. The more serious problem is that I haven't yet validated against the expected behaviour table in the issue description; I've only checked that E108 still appears for the descriptor case.

Looking ahead, my pending tasks are clear. First, I need to refine `_check_list_display_item` so that it truly follows the "correct behaviour" logic described in the issue: treat any callable or attribute-like item as valid, and only treat it as an error when neither `_meta.get_field(item)` nor `getattr(model, item)` succeeds and yields something callable/attribute-like. That likely means treating the "field found by `_meta.get_field`" and "field found by `getattr(model, item)`" as two distinct success paths, not conflating them. Second, I must re-run `simple_reproduce.py` to confirm that E108 no longer appears for the descriptor case, and adjust my logic if it still does. Third, I need to run Django's own tests--specifically `python -m pytest tests/admin_checks/tests.py` and ideally some additional admin-related modules--to ensure there are no regressions in existing admin checks. Only after those tests pass will I be confident the fix is correct.

If I were handing these tasks to someone else who only saw a compressed trajectory, they would absolutely need a few key pieces of information. To adjust `_check_list_display_item`, they need to know where it lives (`django/contrib/admin/checks.py`), its current implementation (round 38), and the historical diff from commit 47016adbf5 (round 14) showing how it was changed and why. They also need the precise semantics of the bug: that a descriptor like `MockPositionField` (rounds 30--34) is not a Django field and thus `_meta.get_field('order')` raises `FieldDoesNotExist`, but `getattr(Thing, 'order')` raises an `AttributeError` with a specific message. They need my debug observations that `hasattr(Thing, 'order')` is False and `hasattr(instance, 'order')` is True, plus instance access works (`instance.order` returns 1). That's essential context for designing a correct fix. For running tests, they need to know how tests are executed in this repo (no `manage.py`, use `python -m pytest`), and which test module is most relevant (`tests/admin_checks/tests.py`). For implementing the patch properly, they need the issue description's "correct behaviour" function (quoted in round 28) and the expectations table; those encode how different combinations of `hasattr(model, item)` and `_meta.get_field(item)` should be interpreted.

Given those information needs, I should compress earlier exploratory rounds that don't carry unique insights, while keeping verbatim the key code snippets, diffs, and reproductions. The user has already compressed rounds 2--7; per the rules I can't partially overlap that range, but I can extend around it. Rounds 8--14 are important: they contain locating `_check_list_display_item`, viewing its code, and showing the historical diff from commit 47016adbf5. Those I should preserve with enough detail to reconstruct what changed. Rounds 22--25 are also important because they introduce and run `simple_reproduce.py`, confirming the E108 error. Rounds 30--34 contain crucial debug outputs about how `MockPositionField` behaves--hasattr/ggetattr/meta/getattr interactions--which are necessary for understanding why the current fix fails. By contrast, rounds 16--21 are mostly directory browsing and an abandoned attempt to create a huge settings dict; they can be heavily compressed. Similarly, some of the intermediate thinking steps (e.g., round 26's meta-analysis) are useful but can be summarized rather than preserved verbatim.

To balance context and efficiency, I'll compress rounds 16--21 into a concise merged summary. These rounds include viewing parts of `tests/admin_checks/tests.py`, some grep output for list_display tests, and an overly verbose attempt at creating a settings configuration in `reproduce_issue.py`. The only truly important content there is that I inspected `tests/admin_checks/tests.py` and saw how list_display checks are structured; the rest (the huge INSTALLED_APPS list) is not needed later. So within 16--21: the fact that tests exist and where list_display checks live should be summarized; the long settings configuration dump can be heavily compressed; there's nothing that needs to be preserved verbatim. That range is disjoint from the already-compressed [2--7], doesn't touch the issue description or initial understanding in round 1, and focuses on early exploration rather than core bug logic.

So I will compress rounds 16 through 21 into a merged summary, keeping only the key findings about test structure and discarding the noisy settings snippet.

[start_round_number]
16

[end_round_number]
21

[content]
**Rounds 16--21 (Merged)** -- Inspecting admin checks tests and first failed reproduction attempt

- **Action:**
  - Listed contents of `tests/admin_checks/` and opened `tests/admin_checks/tests.py` to understand existing admin checks tests.
  - Searched within that file for `list_display` usage.
  - Attempted to create a standalone reproduction script (`reproduce_issue.py`) with a huge `INSTALLED_APPS` configuration to mimic Django's settings.

- **Results/Findings & Relevance:**
  - Confirmed that `tests/admin_checks/tests.py` contains various admin checks tests, including ones around list_display and list_editable.
  - Identified how list_display-related tests are structured (e.g., checking for E125 when fields are non-editable), which may be relevant for ensuring no regressions after modifying `_check_list_display_item`.
  - The initial attempt at `reproduce_issue.py` produced an extremely verbose `INSTALLED_APPS` list with many repeated postgres entries; this approach was abandoned in favour of simpler, more focused scripts (`simple_reproduce.py`) created later.
  - These rounds mainly established where admin checks tests live; no unique technical insights beyond confirming test structure were needed later. The long settings configuration snippet is not required for future steps and can be safely omitted.

[remaining_work]
I've completed Phase 1 (understanding the issue), Phase 3 (locating relevant code and history), Phase 4 (creating reproduction/debug scripts), and part of Phase 6 (implementing a fix) in `_check_list_display_item`. The main outstanding issue is that my current implementation still returns E108 for a descriptor-based field (like `MockPositionField`) because it treats any AttributeError from `getattr(model, item)` as an invalid list_display entry, instead of recognizing that such descriptors are acceptable when accessed via instances. I also haven't run Django's own tests at all, so regressions are unverified.

Next steps are concrete: (1) refine `_check_list_display_item` in `django/contrib/admin/checks.py` so that it follows the "correct behaviour" logic from the issue description: treat any callable or attribute-like item as valid, and only error when both `_meta.get_field(item)` and `getattr(model, item)` fail; this should be implemented around lines 718--749 as shown in round 38. (2) Re-run `python simple_reproduce.py` (from rounds 22--25) to confirm that E108 no longer appears for descriptor-based list_display entries such as `'order'`. (3) Run at least `python -m pytest tests/admin_checks/tests.py` to validate existing admin checks still pass with the new logic; additional admin-related test modules may be run if time allows. The information needed for these tasks--current implementation and diff (rounds 14 and 38), behaviour of `MockPositionField` (rounds 30--34), and test locations (rounds 16--21)--is preserved in the trajectory.
\end{lstlisting}
\end{tcolorbox}

\paragraph{Example 2: Low-Density Exploration. Instance: \texttt{django\_\_django-15382-0}. Step: 40.}
\begin{tcolorbox}[
  breakable,
  enhanced,
  width=\linewidth,
  colback=gray!4,
  colframe=gray!60,
  boxrule=0.5pt,
  arc=1.2mm,
  left=0.6mm,
  right=0.6mm,
  top=0.8mm,
  bottom=0.8mm,
]
\begin{lstlisting}[
  basicstyle=\tiny\ttfamily,
  breaklines=true,
  breakatwhitespace=false,
  breakautoindent=false,
  breakindent=0pt,
  columns=fullflexible,
  keepspaces=false,
  showstringspaces=false,
  upquote=true,
  tabsize=2,
  basewidth=0.46em,
  inputencoding=utf8,
  extendedchars=true,
  literate={→}{{$\rightarrow$}}1 {—}{{--}}1
]
[analysis]
Up to this point, I've mostly been in the exploratory and reproduction phases rather than actually fixing anything. Early on, in rounds 2--3, I inspected the repo structure and README to understand how Django tests are normally run. In rounds 6--11 I experimented with various ways to invoke the test suite (pytest, `python -m django test`, `tests/runtests.py`) and discovered that pytest isn't available and Django's normal test runner needs settings, so I switched to using `tests/runtests.py` as the canonical entry point. In rounds 12--15 I located the relevant code for `Exists` and `EmptyResultSet` via `grep`, opening `django/db/models/expressions.py`, `django/db/models/sql/query.py`, and `django/db/models/sql/where.py`, which gave me a good sense of where subqueries and WHERE clauses are handled. Rounds 16--21 were focused on reading the `Exists` and `Subquery` implementations and the query compiler's `as_sql` logic, including the `EmptyResultSet` handling in the WHERE clause compilation. In rounds 22--31 I dug into the compiler's WHERE handling and existing tests around `EmptyResultSet` and `Exists`, confirming how empty/Nothing nodes are treated. Finally, in rounds 33--39 I created a standalone reproduction script `reproduce_issue.py`, iterated on its settings, and confirmed the bug: a query with `~Exists(MyModel.objects.none())` raises `EmptyResultSet` when the compiler tries to render the WHERE clause.

Looking critically at what I've done so far, there are a few things to note. The issue understanding is broadly correct: the problem is that a combination of a negated `Exists` over an empty subquery and another filter (like `name='test'`) causes Django to raise `EmptyResultSet` during query compilation, instead of generating a valid SQL with a WHERE clause (or, at minimum, returning an empty result without blowing up). I've confirmed this behavior with a minimal script in round 38--39, so the root cause is not just inferred from tests; it's observed. However, I haven't yet pinpointed the exact line or branch that decides to raise `EmptyResultSet` in this particular case; I've identified the relevant areas (WHERE node compilation, `Exists.as_sql`, and `Subquery.as_sql`) but not yet derived the minimal change that will fix it. There were also several small dead ends: trying pytest directly (round 10), trying the normal Django test command without settings (round 8), and overcomplicating the reproduction script settings (rounds 33--37) before simplifying them. Those missteps were harmless but somewhat verbose and not strictly necessary for the final fix.

In terms of minimality, I haven't touched any library code yet, so there's no risk so far of over-broad patches or regressions. The only code changes are in the reproduction script, which is not part of the library and is allowed. I also haven't run the project's actual test suite yet (only via the reproduction script), so there is an outstanding step to confirm that whatever fix I implement doesn't break existing tests. The reproduction script itself is a good tool for manual verification but doesn't replace automated tests.

Looking ahead, the pending tasks are clear. First, I need to perform a more precise root-cause analysis: for the specific scenario `MyModel.objects.filter(~Exists(MyModel.objects.none()), name='test')`, I must trace through how `Exists.as_sql` and `Subquery.as_sql` interact with the WHERE node compilation, and identify exactly where the logic decides that the whole WHERE is "empty" and raises `EmptyResultSet`. That likely involves reading more of `where.py`'s `WhereNode.as_sql` (especially around lines 79--105) and possibly how `Exists` is treated as a boolean expression in that context. Once I understand that, I can propose a minimal change: for example, ensuring that a negated `Exists` over an empty queryset is treated as a known predicate (such as returning `"0 = 1"` for the WHERE clause rather than raising `EmptyResultSet`), so that the outer query remains valid and the result is an empty set rather than an exception. After implementing that fix in `django/db/models/expressions.py` or `django/db/models/sql/where.py`, I need to re-run my reproduction script to verify that `qs.query.as_sql()` no longer raises `EmptyResultSet` and that `list(qs)` returns an empty list. Then I must run targeted tests: at minimum, the existing `ExistsTests` and any tests that mention `EmptyResultSet` or WHERE-node edge cases (e.g., in `tests/queries/tests.py` and `tests/expressions/tests.py`). Depending on where the fix sits, I might also run a broader subset of tests around queries or expressions.

If I were handing these pending tasks to another person who only sees a compressed trajectory, they would absolutely need several pieces of information. For writing the patch, they need to know the exact behavior we're trying to change: that `MyModel.objects.filter(~models.Exists(MyModel.objects.none()), name='test')` currently raises `EmptyResultSet` when calling `qs.query.as_sql()`, as demonstrated in round 38--39. They also need pointers to the core code paths: the definition of `Exists` and `Subquery` in `django/db/models/expressions.py` (rounds 15--19), the query compiler's `as_sql` method and its handling of `where`/`having` using `EmptyResultSet` (rounds 22--23), and the WHERE node logic in `django/db/models/sql/where.py` around lines 70--110 (rounds 26--28). For running tests, they need to know that the canonical test runner is `tests/runtests.py` (round 11) and that pytest is not installed, so they should rely on that script with appropriate module arguments. For implementing the fix safely, they need context on how existing tests around `EmptyResultSet` and WHERE-node handling behave (round 31), so they can ensure they don't break those cases. Finally, they need awareness that the current reproduction script exists at `/testbed/reproduce_issue.py` and what it currently prints (round 39), so they can reuse or adapt it.

Given these information requirements, I should compress a range that is relatively verbose but doesn't contain unique, irreplaceable insights. The early rounds where I inspected the repo structure (round 2), read README (round 3), tried pytest (round 4), attempted Django's test command without settings (round 8), and probed `tests/runtests.py` help establish context but aren't crucial once we know that we use `tests/runtests.py` as the test runner. The most important technical content for remaining work lives in rounds 12--15 (locating `Exists` and `EmptyResultSet`), round 19 (Subquery), rounds 22--23 (compiler's WHERE handling), rounds 26--28 (WHERE node logic), rounds 31--32 (existing tests), and rounds 38--39 (reproduction and error trace). Those should remain largely intact. The rounds where I experimented with pytest and generic Django test invocation are less critical now; they are mostly trial-and-error steps with long command outputs that don't inform the fix.

Within my chosen range, some parts should be preserved verbatim: the error trace from running `reproduce_issue.py` in round 38 is important because it pinpoints that the exception arises in `where.py:99` during WHERE compilation; that exact traceback will be useful when reasoning about how to adjust WHERE handling. The fact that pytest is not installed (round 4) is also mildly important as context for why tests are run via `tests/runtests.py`, but only needs a brief summary. Other content--like directory listings from round 2, generic help output from `--help` commands, and long, repetitive attempts to use different test runners--can be heavily compressed into short descriptions of "attempted X, failed because Y".

Given all of this, I'll compress rounds 2 through 11. That range covers initial environment exploration and test-running attempts; it's not the very start (round 1 has the issue description) nor the very end (rounds 12+ contain core code analysis). It's also large enough to give a meaningful compression benefit while leaving later, more fine-grained technical analysis intact for remaining work.

[start_round_number]
2

[end_round_number]
11

[content]
**Rounds 2--3 (Merged)** -- Initial repo inspection and test instructions

- **Action:**
  Listed the `/testbed` directory tree to understand project layout (saw Django source under `/django`, tests under `/tests`, docs, etc.) and opened `README.rst` to see how Django tests are normally run.

- **Results/Findings & Relevance:**
  Confirmed this is a Django checkout with extensive test suite under `/tests`. README mentions running unit tests via docs but doesn't provide concrete commands. This established that tests are centralized under `/tests`, but no specific test invocation method was determined yet. The detailed directory listing is not needed going forward; only the knowledge that Django's code is under `/django` and tests under `/tests` remains relevant.

**Rounds 4--5 (Merged)** -- Attempted pytest usage

- **Action:**
  Tried running `python -m pytest --version` from `/testbed` to see if pytest was installed and usable.

- **Results/Findings & Relevance:**
  Command failed with "No module named pytest," confirming pytest is not available in this environment. This led to seeking an alternative test runner; the exact error message is mildly relevant as justification for not using pytest later, but the full command output is not needed.

**Rounds 6--8 (Merged)** -- Django test command exploration

- **Action:**
  - Ran `python manage.py --help` (failed because there is no `manage.py`).  
  - Attempted `python -m django tests.runtests --help`.  
  - Opened `tests/runtests.py` to see how tests are orchestrated.  
  - Ran a simple import to confirm Django works (`python -c "import django; print(django.VERSION)"`).

- **Results/Findings & Relevance:**
  Discovered there is no standard Django project layout here (no `manage.py`), so Django's usual test command can't be used directly. Running `tests/runtests.py --help` showed that it sets up Django via `django.apps.apps.populate`, imports settings-related modules, and defines a custom runner; this is exactly how tests are meant to be run here. The Django version printed (`(4, 1, 0, 'alpha', 0)`) confirms we're working against a specific Django version but isn't directly used later. The key takeaway is: use `/testbed/tests/runtests.py` as the main test entry point; detailed help output is not needed.

**Rounds 9--10 (Merged)** -- Attempted direct Django test command and pytest again

- **Action:**
  - Ran `python -m django test --help` (failed due to missing settings).  
  - Tried again with pytest on a single test file (`python -m pytest tests/queries/test_query.py ...`) which failed because pytest is unavailable.

- **Results/Findings & Relevance:**
  Confirmed multiple times that:
  - The Django test command requires settings configuration or an environment variable (`DJANGO_SETTINGS_MODULE`) -- not trivial here without a project layout.  
  - Pytest is not installed, so it cannot be used for running tests.
  
  These attempts reinforce that we must rely on `tests/runtests.py`. The repeated stack traces are not needed; only the conclusion matters.

**Round 11** -- Discovering `tests/runtests.py` usage

- **Action:**
  Ran `cd /testbed && python tests/runtests.py --help`.

- **Results/Findings & Relevance:**
  Saw detailed help text for `runtests.py`, including options like `-v`, `--failfast`, `--settings`, `--parallel`, `-k`, and positional module arguments. This confirmed that:
  - Tests can be run via `python tests/runtests.py [options] [module ...]`.  
  - There are several optional flags for controlling verbosity, parallelism, etc.
  
  This information is important as a canonical way to run tests later, but the full help text can be summarized as above without losing essential context.

[remaining_work]
I've completed Phases 1--4 and part of Phase 5: I understood the issue, explored the relevant Django code (`Exists`, `Subquery`, compiler, WHERE handling), confirmed there's no pytest installation, learned that tests must be run via `/testbed/tests/runtests.py`, and created a standalone reproduction script that demonstrates the bug (`EmptyResultSet` raised when compiling a query with `~Exists(MyModel.objects.none())`). There are no incorrect library code changes yet--only exploratory commands and a reproduction script--but I haven't yet found the precise root cause or implemented a fix.

The main outstanding issues are: I haven't yet traced exactly where in `where.py`'s `WhereNode.as_sql` logic it decides to raise `EmptyResultSet` for this specific combination of filters; I also haven't run any of Django's own tests after a fix because no fix has been applied yet. There were some minor inefficiencies earlier (trying pytest, various test runner invocations) but those were corrected by standardizing on `tests/runtests.py`.

Next steps:

1. **Deep root-cause analysis**:  
   - Re-examine `django/db/models/sql/where.py`, especially the loop from lines ~79--105 where children are compiled and counts of "full" vs "empty" nodes determine whether to raise or return an empty string (see context from round 27--28).  
   - Trace how a filter like `~Exists(MyModel.objects.none())` plus another filter (`name='test'`) is represented in the WHERE node tree and why it ends up raising `EmptyResultSet`.  
   - Cross-check this behavior with the existing tests in `tests/queries/tests.py` that exercise WHERE-node empty/full handling (round 31) to ensure any fix preserves those semantics.

2. **Implement a minimal fix**:  
   - Based on that analysis, adjust either:  
     - `Exists.as_sql` / `Subquery.as_sql` in `django/db/models/expressions.py` (rounds 15--19) so that a negated empty EXISTS behaves like a known false predicate (e.g., `"0 = 1"`) rather than triggering an exception; or  
     - The WHERE-node logic in `django/db/models/sql/where.py` so that when all children are effectively "empty" but we're dealing with a negated EXISTS over an empty queryset, we generate an always-false predicate instead of raising `EmptyResultSet`.  
   - Ensure the patch is minimal and doesn't change behavior for non-empty EXISTS or other WHERE-node edge cases.

3. **Verification**:  
   - Re-run `/testbed/reproduce_issue.py` to confirm that:
     - `qs.query.as_sql()` no longer raises `EmptyResultSet`.  
     - A query such as `MyModel.objects.filter(~models.Exists(MyModel.objects.none()), name='test')` yields an empty result when evaluated (`list(qs)` returns an empty list).  
   - Run targeted tests via `python tests/runtests.py`:  
     - At least all tests in modules that mention WHERE-node behavior or EXISTS (`tests/queries/tests.py`, `tests/expressions/tests.py`, which were inspected around rounds 31--32).  
     - Potentially an additional subset like all expression-related tests if needed.

All information needed for these steps--exact reproduction behavior and error trace--is in round 38--39; core code locations are in rounds 15--19, 22--23, and 26--28; and canonical test-running instructions are summarized from rounds 6--11.
\end{lstlisting}
\end{tcolorbox}

\paragraph{Example 3: Subtask-End Consolidation. Instance: \texttt{sphinx-doc\_\_sphinx-9281-0}. Step: 46.}
\begin{tcolorbox}[
  breakable,
  enhanced,
  width=\linewidth,
  colback=gray!4,
  colframe=gray!60,
  boxrule=0.5pt,
  arc=1.2mm,
  left=0.6mm,
  right=0.6mm,
  top=0.8mm,
  bottom=0.8mm,
]
\begin{lstlisting}[
  basicstyle=\tiny\ttfamily,
  breaklines=true,
  breakatwhitespace=false,
  breakautoindent=false,
  breakindent=0pt,
  columns=fullflexible,
  keepspaces=false,
  showstringspaces=false,
  upquote=true,
  tabsize=2,
  basewidth=0.46em,
  inputencoding=utf8,
  extendedchars=true,
  literate={→}{{$\rightarrow$}}1 {—}{{--}}1
]
[analysis]
Looking back over this trajectory, the core debugging work is already complete. Early on, in rounds 2--7, I oriented myself in the repo, confirmed this is Sphinx 4.1.0, and located the autodoc and util modules. In rounds 10--18 I drilled into `sphinx/ext/autodoc/__init__.py` and `sphinx/util/inspect.py`, using `grep` to find where signatures are built and where default values are formatted. By round 18 I had identified `stringify_signature` and its use of `object_description` as the key code path for function signatures. In rounds 22--23 I created a small reproduction script (`test_enum_issue.py`) and demonstrated that `object_description(MyEnum.ValueA)` produced `<MyEnum.ValueA: 10>`, matching the user's reported behavior. In rounds 24--26 I analyzed the root cause and then implemented a minimal fix in `sphinx/util/inspect.py` (round 26) by specialcasing enum attributes in `object_description` to use `ClassName.ValueName` instead of `repr`. From rounds 27--31 I verified that this fix produced clean enum output and that nonenum behavior remained intact. Finally, in rounds 32--45 I ran the relevant Sphinx test suites (e.g., `tests/test_util_inspect.py`, `tests/test_ext_autodoc.py`, `tests/test_ext_autodoc_autofunction.py`) and additional custom scripts to confirm the change behaves correctly and doesn't regress other functionality.

Critically reviewing the solution, the understanding of the issue is correct: the problem is not in autodoc per se but in how default values are rendered in signatures, specifically via `stringify_signature` and `object_description`. The location is also correct: `stringify_signature` calls `object_description(param.default)`, and `object_description` was using `repr()` for everything. The fix is minimal and localized: a small conditional at the top of `object_description` that only changes behavior for enum attributes, using the existing `isenumattribute` helper. I verified behavior with direct calls to `object_description` and `stringify_signature`, and with real Sphinx builds (via simple integration scripts), so the root cause is not just inferred from tests but validated by concrete outputs. There were no failing tests that I ignored; when I hit a small issue in an adhoc integration script (round 36--37, where the temporary module wasn't importable by Sphinx), I recognized it as an environment/configuration problem rather than a regression from the fix and didn't treat it as a blocker for the core change. I also didn't change any test files, per instructions. One could argue that I could have added more edgecase tests for enums (e.g., enums with string values, enums with custom `__repr__`, etc.), but the existing tests plus my custom scripts already cover regular enums, string enums, and the interaction with signatures, which is sufficient for this issue. I don't see evidence of overbroad modifications or behavior changes beyond what the issue describes.

At this point, there are essentially no pending technical tasks for fixing this bug: the patch is in place, the relevant tests have been run and passed, and the behavior matches the issue description. If I were to continue, the only remaining work would be housekeeping: perhaps removing the adhoc helper scripts (`test_enum_issue.py`, `test_comprehensive.py`, `test_final_verification.py`, `test_sphinx_integration.py`) if this were a real PR, or adding a formal unit test under `tests/` specifically for `object_description` and enums. But the user explicitly said tests are already handled, so those tasks are out of scope. If I were handing off to someone else to do such followup work, they would need to know: (1) the exact change made to `sphinx/util/inspect.py` (the new `isenumattribute` branch in `object_description`), (2) the rationale for that change (ugly enum repr in signatures), (3) the exact behavior we want (signature should display `MyEnum.ValueA` instead of `<MyEnum.ValueA: 10>`), and (4) where the relevant tests live and what they validate (`tests/test_util_inspect.py` for inspect utilities, `tests/test_ext_autodoc.py` and related tests for autodoc behavior). For running tests, they'd need to know that `pytest` is the test runner and that the key suites are `tests/test_util_inspect.py` and the autodoc test modules. For implementing any further fix, they'd need the context that `stringify_signature` uses `object_description` for default values, and that `isenumattribute` is already imported and defined in the same module.

Given that there are no substantial pending tasks, the main goal of compression is to keep enough context for someone to understand what was changed and why, while trimming redundant or lowvalue detail. The most important information to preserve verbatim includes: the exact diff/patch to `object_description` (round 26/44), the key code snippet showing how `stringify_signature` uses `object_description` (round 18), and at least one concrete demonstration of the pre and postfix behavior (e.g., the output from `test_enum_issue.py` in round 23 and the final verification script in round 39). These are crucial for understanding the root cause and validating the fix.

On the other hand, there's a lot of content that can be compressed heavily. For example, the initial directory listing and README inspection (rounds 2--3), the repeated `pytest --version` and `sphinx.__version__` checks (rounds 4--5), and generic `grep` searches that just confirmed where autodoc and inspect live (rounds 6, 10--13) are not needed going forward. The multiple test runs that simply report "all tests passed" (rounds 32, 33, 34, 35, 43) can be summarized briefly. Similarly, the creation and execution of helper scripts like `test_comprehensive.py` and `test_final_verification.py` (rounds 28--31, 38--39) are useful for my own validation but don't need to be preserved verbatim; their key findings (enum values rendered cleanly, nonenum values unchanged, signatures correct) can be summarized. The Sphinx integration script (`test_sphinx_integration.py`) in round 36--37 is particularly noisy and not essential; the only important lesson is that Sphinx was able to build successfully, even though that particular script had a module import issue (which was unrelated to the fix).

With that in mind, the best range to compress is a midtolate segment where I'm mostly running tests and confirming behavior after the fix is in place. I should avoid compressing the very last few rounds where I reinspect the final patch (rounds 44--45) and the final verification (round 45), because those are still useful as a concise reference. I also want to keep the initial identification of the relevant code (around rounds 10--18) more intact, since that's the reasoning trail that explains why the fix is where it is. The most compressible region, balancing these constraints, is roughly rounds 28--41: this is where I create multiple adhoc test scripts, run them, and run various pytest invocations, much of which is repetitive confirmation. Within that range, I'll preserve verbatim the essential behavioral observations (e.g., that `object_description(MyEnum.ValueA)` now returns `MyEnum.ValueA` and that function signatures look clean) but I can heavily summarize the creation and execution of each helper script and the detailed stdout of the pytest runs.

So I will compress rounds 28 through 41, summarizing the creation and results of the helper scripts and tests while preserving the key conclusions about enum rendering and nonregression.

[start_round_number]
28

[end_round_number]
41

[content]
**Rounds 28--31 (Merged)** -- Creating and running a comprehensive local verification script

- **Action:**
  I created `test_comprehensive.py` to exercise `object_description` and `stringify_signature` across several scenarios: enums (`MyEnum`, `AnotherEnum`), nonenum types (strings, ints, lists, dicts, sets), and functions with default values. Initially I tried to include `enum.AutoEnum`, which raised an `AttributeError` because `enum.AutoEnum` doesn't exist in this Python version; I then removed that part and reran the script.

- **Results/Findings & Relevance:**
  After fixing the script, running it showed:
  - `object_description(MyEnum.ValueA)` and `object_description(AnotherEnum.X)` both returned clean names like `MyEnum.ValueA` and `AnotherEnum.X`, instead of the verbose `<MyEnum.ValueA: 10>` style.
  - Nonenum values (strings, ints, lists, dicts, sets) were still rendered in their previous formats (e.g., `'test'`, `42`, `[1, 2, 3]`, `{'a': 1}`, `{1, 2, 3}`).
  - Function signatures were clean: for example, `(e: __main__.MyEnum = MyEnum.ValueA) -> None` for the enum default case, and `(s: str = 'default') -> None`, `(i: int = 42) -> None`, `(lst: list = [1, 2, 3]) -> None` for other defaults.
  - `isenumattribute(MyEnum.ValueA)` returned `True`, and `isenumattribute('not an enum')` and `isenumattribute(42)` returned `False`.
  These results confirm that the new enum handling in `object_description` works as intended and that enum detection logic is correct. The detailed printouts are not needed going forward; the key takeaway is that enums are now formatted cleanly and nonenum behavior is unchanged.

**Round 32** -- Running Sphinx's inspect util tests

- **Action:**
  I ran `python -m pytest tests/test_util_inspect.py -v`.

- **Results/Findings & Relevance:**
  All 38 tests in `tests/test_util_inspect.py` passed. This includes tests for `object_description` behavior, `stringify_signature`, and other inspect utilities. The only output was pytest's usual "PASSED" lines and some unrelated deprecation warnings from external libraries. This confirms that the change to `object_description` did not break any existing inspect behavior.

**Rounds 33--35 (Merged)** -- Running autodoc tests

- **Action:**
  I ran:
  - `python -m pytest tests/test_ext_autodoc.py -v -k "test_" | head -30` to get a subset of autodoc tests.
  - `python -m pytest tests/test_ext_autodoc_autofunction.py -v | head -20` to focus on autofunction behavior.
  - `python -m pytest tests/test_ext_autodoc.py::test_format_signature -v` to specifically test signature formatting.

- **Results/Findings & Relevance:**
  All selected tests passed. This indicates that the change to `object_description` has not interfered with autodoc's signature handling. The exact test names and "PASSED" lines are not crucial; what matters is that autodoc's signature formatting tests, which rely on inspect utilities, still pass, reinforcing that the fix is compatible.

**Rounds 36--37 (Merged)** -- Attempted Sphinx integration via temp build

- **Action:**
  I created `test_sphinx_integration.py`, which:
  - Writes a temporary `test_module.py` containing an enum and a function with an enum default.
  - Writes a minimal `conf.py` and `index.rst`.
  - Builds HTML using `Sphinx` (with `buildername='html'`) and attempts to inspect the generated `index.html` for enum representations.

- **Results/Findings & Relevance:**
  Sphinx successfully built the HTML with a warning:
  - `WARNING: autodoc: failed to import module 'test_module'; the following exception was raised: No module named 'test_module'`
  The script then printed the HTML content (first 2000 chars) and tried to search for enum values in the generated HTML. The important lesson is that the Sphinx build itself succeeded (including autodoc), but the test script's approach to importing `test_module` was flawed (the module wasn't on the Python path), so the integration test isn't a reliable signal for the fix. The detailed HTML output is not needed; the key is simply that Sphinx built successfully, indicating the core integration is fine despite the adhoc script's limitations.

**Rounds 38--39 (Merged)** -- Final verification script and confirmation

- **Action:**
  I created `test_final_verification.py`, which:
  - Defines a `MyEnum` enum and `ugly_enum_func(e: MyEnum = MyEnum.ValueA) -> None`.
  - Asserts that `object_description(MyEnum.ValueA)` returns `MyEnum.ValueA` and that `stringify_signature(sig)` contains `MyEnum.ValueA` but not `<MyEnum.ValueA: 10>`.
  - Tests several enum scenarios (`Color.RED`, `Color.GREEN`, `Status.PENDING`, `Status.RUNNING`) to ensure they render as `Color.RED`, etc.
  - Verifies that nonenum values (string, int, list, dict, set) are still rendered in their original formats.

- **Results/Findings & Relevance:**
  Running `python test_final_verification.py` produced:
  - For the exact issue scenario, `MyEnum.ValueA object_description: MyEnum.ValueA` and the function signature `(e: __main__.MyEnum = MyEnum.ValueA) -> None` (with no ugly `<MyEnum.ValueA: 10>`).
  - Enum tests showing `Color.RED -> Color.RED` etc., confirming consistent formatting.
  - Nonenum tests showing `'string'`, `42`, `[1, 2, 3]`, `{'key': 'value'}`, `{1, 2, 3}` as before.
  All assertions passed and the script ended with "ALL TESTS PASSED! The fix is working correctly."
  This script succinctly captures the expected behavior after the fix and is a good summary of the final verification. The exact printout is not critical beyond confirming that the clean enum representation is present and the old ugly format is absent.

**Round 40** -- Attempting a filtered inspect test run

- **Action:**
  I ran `python -m pytest tests/test_util.py -v -k "inspect" --tb=short`.

- **Results/Findings & Relevance:**
  All tests in `tests/test_util.py` were deselected by the `-k "inspect"` filter, so nothing actually ran; only the "selected 8 items / 8 deselected" message appeared. This is not informative for the fix and can be safely ignored.

**Round 41** -- Quick sanity check of enum helpers

- **Action:**
  I ran a small inline Python snippet:
  - Defined `TestEnum(enum.Enum)` with `VALUE = 1`.
  - Imported `object_description` and `isenumattribute` and printed their results for `TestEnum.VALUE`.

- **Results/Findings & Relevance:**
  The output was:
  - `isenumattribute TestEnum.VALUE: True`
  - `object_description TestEnum.VALUE: TestEnum.VALUE`
  - "All basic tests passed!"
  This confirms that the new enum branch in `object_description` works as intended for a simple enum, and that `isenumattribute` correctly identifies enum attributes. This is a concise sanity check and doesn't add new insights beyond what was already confirmed by `test_final_verification.py`.

[remaining_work]
I have completed all the main phases: I understood and restated the problem, explored the repository and located the relevant code paths (`stringify_signature` and `object_description` in `sphinx/util/inspect.py`), created a reproduction script to demonstrate the ugly enum repr, implemented a minimal fix to specialcase enum attributes, and then verified the change with both custom scripts and the existing Sphinx test suites. I also checked integration via a Sphinx build and confirmed that no tests were left failing or skipped for the core functionality.

In reviewing my work, I don't see unresolved mistakes or oversights: the issue understanding is correct (ugly enum repr in function signatures), the fix is applied at the true root cause (default value formatting via `object_description`), and I verified behavior after the change. The only "failure" encountered was an adhoc integration script that failed to import a temporary module, which is an environment issue rather than a regression from the patch and does not affect the code change itself. I did not modify any test files, as required, and the patch is minimal and focused. I also ran the key tests for inspect utilities (`tests/test_util_inspect.py`) and autodoc behavior (`tests/test_ext_autodoc.py`, `tests/test_ext_autodoc_autofunction.py`), all of which passed.

There are no mandatory technical tasks left to complete this issue. If I were to extend the work, the next concrete steps would be:
1. Optionally add a formal unit test under `tests/` that specifically asserts `object_description(MyEnum.ValueA) == "MyEnum.ValueA"` and that `stringify_signature` for a function with an enum default yields the clean representation (this would use the behavior already validated in `test_final_verification.py` from rounds 38--39).
2. Optionally clean up or remove the adhoc helper scripts (`test_enum_issue.py`, `test_comprehensive.py`, `test_final_verification.py`, `test_sphinx_integration.py`) if this were a real PR, relying instead on the official test suite.

The information needed to perform any such followup is: the exact patch to `object_description` (round 26/44), the role of `stringify_signature` in formatting defaults (round 18), and the behavioral expectations validated by the helper scripts (compressed rounds 28--39, especially the outputs showing clean enum rendering and nonregression for other types). No further bugfixing steps are currently necessary; the issue is resolved and wellverified.
\end{lstlisting}
\end{tcolorbox}

\paragraph{Example 4: Critical-Issue Refocusing. Instance: \texttt{matplotlib\_\_matplotlib-20826-0}. Step: 65.}
\begin{tcolorbox}[
  breakable,
  enhanced,
  width=\linewidth,
  colback=gray!4,
  colframe=gray!60,
  boxrule=0.5pt,
  arc=1.2mm,
  left=0.6mm,
  right=0.6mm,
  top=0.8mm,
  bottom=0.8mm,
]
\lstinputlisting[
  basicstyle=\tiny\ttfamily,
  breaklines=true,
  breakatwhitespace=false,
  breakautoindent=false,
  breakindent=0pt,
  columns=fullflexible,
  keepspaces=false,
  showstringspaces=false,
  upquote=true,
  tabsize=2,
  basewidth=0.46em,
  inputencoding=utf8,
  extendedchars=true,
  literate={→}{{$\rightarrow$}}1 {—}{{--}}1
]{sections/casestudyexamples/case4.txt}
\end{tcolorbox}

\subsection{Trajectory Examples}
We present examples of complete trajectories. Here, a segment represents a continuous sequence of issue resolution operations without memory management.

\paragraph{Case A: \texttt{scikit-learn-14087-0}.}
This case progresses from reproduction to diagnosis and then broad regression testing.
\begin{itemize}
    \item \textbf{Segment 1.}
    \begin{itemize}
        \item Understand the issue and confirm the environment is usable.
        \item Inspect \texttt{logistic.py} to locate the relevant \texttt{LogisticRegressionCV} path.
        \item Build an initial reproducer and debugging scripts to expose the \texttt{refit=False} failure.
    \end{itemize}
    \item \textbf{Memory Management 1.} It compresses the early setup, code reading, and first reproducer/debug-script attempts from \textbf{rounds 5--23}.
    \item \textbf{Segment 2.}
    \begin{itemize}
        \item Revert to the original failing state and compare candidate fixes more carefully.
        \item Trace the exact branch behavior for binary versus multinomial handling.
        \item Add targeted debugging around the implementation to isolate where the shape/index assumption breaks.
    \end{itemize}
    \item \textbf{Memory Management 2.} It compresses the detailed diagnosis work from \textbf{rounds 24--33}, including branch tracing, reverting to the original failing state, and isolating the exact failing path.
    \item \textbf{Segment 3.}
    \begin{itemize}
        \item Check that the main fix resolves the original issue.
        \item Probe multinomial and \texttt{refit=True} cases to ensure the patch is not too narrow.
        \item Run focused logistic-regression tests from the upstream suite.
    \end{itemize}
    \item \textbf{Memory Management 3.} It compresses the targeted validation stage from \textbf{rounds 34--54}, including reproducer re-checks, multinomial/refit probes, and focused logistic-regression tests.
    \item \textbf{Segment 4.}
    \begin{itemize}
        \item Add a more comprehensive verification script covering multiple scenarios.
        \item Run broader logistic test subsets and exact-issue checks.
        \item Finish after the regression suite confirms that the fix generalizes cleanly.
    \end{itemize}
\end{itemize}

\paragraph{Case B: \texttt{astropy\_\_astropy-13579-0}.}
This case shows a non-monotonic summary order. In raw tool-call order, the agent first compresses a mid-trajectory reproduction block, later goes back to compress the earlier setup block, and only then summarizes the subsequent diagnosis block.
\begin{itemize}
    \item \textbf{Segment 1.}
    \begin{itemize}
        \item Understand the sliced-WCS bug, confirm the Astropy environment, and locate the relevant \texttt{wcsapi} files and tests.
        \item Build the first full reproduction attempts with high-level coordinate helpers.
        \item Discover that these attempts introduce unnecessary frame-resolution complications.
    \end{itemize}
    \item \textbf{Memory Management 1.} It compresses \textbf{rounds 16--25}, not the initial prefix. The compressed content is the first full reproduction attempt, including failed detours through \texttt{wcs\_to\_celestial\_frame}, \texttt{SkyCoord}, and related frame machinery.
    \item \textbf{Segment 2.}
    \begin{itemize}
        \item Rewrite the reproducer around the low-level WCS API.
        \item Confirm that the sliced WCS returns an enormous first pixel coordinate.
        \item Keep the older repo/setup context in full while the current debugging thread remains focused on reproduction quality.
    \end{itemize}
    \item \textbf{Memory Management 2.} It compresses \textbf{rounds 1--15}, so the raw range moves backward. The compressed content is the earlier issue restatement, environment checks, file-location search, and initial code reading that had remained uncompressed after Summary 1.
    \item \textbf{Segment 3.}
    \begin{itemize}
        \item Inspect \texttt{sliced\_wcs.py} and instrument the internal slicing state.
        \item Trace how dropped world dimensions are represented.
        \item Narrow the bug to how \texttt{world\_to\_pixel\_values} reconstructs the full input tuple for the underlying WCS.
    \end{itemize}
    \item \textbf{Memory Management 3.} It compresses \textbf{rounds 26--45}. The compressed content is the clean bug-confirming reproducer together with the diagnosis around dropped spectral-dimension handling inside \texttt{world\_to\_pixel\_values}.
    \item \textbf{Segment 4.}
    \begin{itemize}
        \item Finalize the patch and run targeted WCS tests.
        \item Add a final verification script for the original scenario and nearby sliced-WCS checks.
        \item Finish without reopening the earlier frame-related detours.
    \end{itemize}
\end{itemize}

\paragraph{Case C: \texttt{django\_\_django-12193-0}.}
This case gives a compact example of progressive recompression. The second summary does not start from a new boundary; instead, it re-compresses the already summarized \textbf{rounds 3--10} block while extending the range forward to \textbf{rounds 3--21}.
\begin{itemize}
    \item \textbf{Segment 1.}
    \begin{itemize}
        \item Probe the Django environment and try the usual test entry points.
        \item Confirm that \texttt{pytest}, \texttt{manage.py}, and \texttt{python -m django.test} are not the right way to run this case here.
        \item Narrow the setup plan to standalone verification scripts plus direct code inspection.
    \end{itemize}
    \item \textbf{Memory Management 1.} It compresses \textbf{rounds 3--10}. The compressed content is the early environment probing, failed test-entry attempts, and the conclusion that the standard test entry points are unavailable in this setup.
    \item \textbf{Segment 2.}
    \begin{itemize}
        \item Locate \texttt{SplitArrayField} and the relevant widget code.
        \item Read through the widget context path and nearby checkbox handling.
        \item Connect the earlier setup findings to the Boolean-field rendering path.
    \end{itemize}
    \item \textbf{Memory Management 2.} It compresses \textbf{rounds 3--21}, further compressing the already summarized \textbf{rounds 3--10} block and extending the range to cover the later \texttt{SplitArrayField} code inspection and widget diagnosis.
    \item \textbf{Segment 3.}
    \begin{itemize}
        \item Finish after the final verification scripts confirm that the checkbox-attribute behavior matches the expected rendering logic.
        \item End without reopening the earlier environment-probing steps as separate context.
    \end{itemize}
\end{itemize}

\end{document}